\documentclass[11pt]{article}
\usepackage{jheppub}
\bibstyle{JHEP}

\usepackage[english]{babel} 
\usepackage[utf8]{inputenc}  
\usepackage[T1]{fontenc}
\usepackage{lmodern}
\usepackage{amsmath}         
\usepackage{amsfonts}          
\usepackage{amssymb}
\usepackage{graphicx} 
\usepackage{comment}
\usepackage{bbold}
\usepackage{bbm}
\usepackage[colorlinks=true,urlcolor=blue,linkcolor=blue,citecolor=blue,linktocpage=true]{hyperref}
\usepackage[colorinlistoftodos]{todonotes} 
\usepackage{nicefrac} 
\usepackage{subcaption}

\newcommand{\NN}{{\mathbb{N}}}
\newcommand{\RR}{{\mathbb{R}}}
\newcommand{\ZZ}{{\mathbb{Z}}}
\newcommand{\Tr}{{\mathrm{Tr}}}
\newcommand{\cH}{{\mathcal{H}}}
\newcommand{\cO}{{\mathcal{O}}}

\newcommand{\cZ}{{\mathcal{Z}}}

\newcommand{\ee}{{\mathrm e}}
\newcommand{\ii}{{\mathrm i}}
\newcommand{\CC}{{\mathbb{C}}}
\newcommand{\cR}{{\mathcal R}}

\newcommand{\ndif}[1][]{{\mathrm{d}^{#1} \!}}
\newcommand{\od}[3][]{\frac{\ndif[#1]{#2}}{\ndif{#3}^{#1}}}
\newcommand{\dif}[2][]{{\ndif[#1] #2 \,}}

\newcommand{\qtext}[1]{\quad \text{#1} \quad}

\newcommand{\gns}[1][]{{|\Omega_{#1}\rangle}}
\newcommand{\gnsev}[2][]{{\langle \Omega_{#1} | #2 |\Omega_{#1}\rangle}}
\newcommand{\Gmod}{G_{\text{mod}}}

\usepackage{tikz}
\usetikzlibrary{calc,decorations.pathmorphing}

\title{ Resolving modular flow: a toolkit for free fermions }

\author[a]{Johanna Erdmenger,}
\author[a]{Pascal Fries,}
\author[b]{Ignacio A. Reyes}
\author[a]{and Christian P. Simon}

\affiliation[a]{Institute for Theoretical Physics and Astrophysics, Julius-Maximilians-Universit\"at W\"urzburg, Am Hubland, 97074 W\"{u}rzburg, Germany}
\affiliation[b]{Max Planck Institute for Gravitational Physics (Albert Einstein Institute),\\Am M\"uhlenberg 1, 14476 Potsdam-Golm, Germany}

\emailAdd{erdmenger@physik.uni-wuerzburg.de}
\emailAdd{pascal.fries@physik.uni-wuerzburg.de}
\emailAdd{christian.simon@physik.uni-wuerzburg.de}
\emailAdd{ignacio.reyes@aei.mpg.de}

\abstract{Modular flow is a symmetry of the algebra of observables associated to spacetime
  regions. Being closely related to entanglement, it has played a key role in recent
  connections between information theory, QFT and gravity. However, little is known about
  its action beyond highly symmetric cases. The key idea of this work is to introduce a
  new formula for modular flows for free chiral fermions in $1+1$ dimensions, working
  directly from the \textit{resolvent}, a standard technique in complex analysis. We
  present novel results -- not fixed by conformal symmetry -- for disjoint regions on the
  plane, cylinder and torus. Depending on temperature and boundary conditions, these
  display different behaviour ranging from purely local to non-local in relation to the
  mixing of operators at spacelike separation. We find the modular two-point function,
  whose analytic structure is in precise agreement with the KMS condition that governs
  modular evolution. Our ready-to-use formulae may provide new ingredients to explore the
  connection between spacetime and entanglement.}

\begin{document} 
\maketitle

\section{Introduction}

Recent years have witnessed significant progress in our understanding of the role that entanglement, as well as other ideas from quantum information theory, play in the context of high energy physics, including Quantum Field Theory (QFT) and gravity. A remarkable example of this exchange of ideas between different research areas is the Ryu-Takayanagi formula \cite{Ryu:2006bv} for the entanglement entropy in the AdS/CFT correspondence, which generalises the Bekenstein-Hawking area law for the black hole entropy.

One of the key concepts that have aided this progress is that of \textit{modular flow}. Loosely speaking, the modular flow $\sigma_t$  of an operator is given by a generalised time evolution with the density matrix $\rho$  itself, $\sigma_t(\cO):=\rho^{\ii t} \cO \rho^{-\ii t}$.
An important property of this flow is that, when introduced in expectation values, it must satisfy a periodicity condition in imaginary time known as the Kubo-Martin-Schwinger (KMS) condition.  Originally introduced within algebraic QFT~\cite{Takesaki:1970kop,Haag:1992hx,Brattelli:1997fuh,Borchers:2000pv,Takesaki:2003ght}, modular flow and its associated generator, the modular Hamiltonian, have found applications across a wide spectrum of topics due to its close connection to quantum information measures. This includes modular theory \cite{Lashkari:2018nsl,Witten:2018zxz,Lashkari:2019ixo}, relative entropy in QFT~\cite{Sarosi:2017rsq,Casini:2017roe,Blanco:2017akw}, entropy and energy inequalities~\cite{Casini:2008cr,Blanco:2013lea,Faulkner:2016mzt,Balakrishnan:2017bjg,Ceyhan:2018zfg}, conformal field theories \cite{Lashkari:2015dia,Cardy:2016fqc,Lashkari:2018oke,Long:2019fay}, and bulk reconstruction in gauge/gravity duality~\cite{Casini:2011kv,Blanco:2013joa,Jafferis:2014lza,Jafferis:2015del,Lashkari:2013koa,Koeller:2017njr,Czech:2017zfq,Chen:2018rgz,Belin:2018juv,Abt:2018zif,Faulkner:2018faa,Jefferson:2018ksk,Czech:2019vih,deBoer:2019uem,Arias:2020qpg}. For free fermions, the use of the resolvent method was first introduced in \cite{Casini:2009vk} to study the vacuum on the plane, and subsequently for the  cylinder \cite{Klich:2015ina}, and torus \cite{Fries:2019ozf,PhysRevD.100.025003}. The modular two-point function on the cylinder was studied in \cite{Hollands:2019hje}.

Despite the many contexts in which modular flow appears, there are very few cases where its action is explicitly known. 
In the general context of QFT, the vacuum modular flow in a Rindler wedge is fixed by Poincaré symmetry alone\,\cite{Bisognano:1975ih}, while conformal symmetry fixes it for diamond shaped geometries\,\cite{Hislop1982zhn}. Anything beyond these cases, be it the choice of another state or a different region, depends on the details of the theory under consideration and is largely unknown. 
Many discussions are concerned with universal properties of modular flows, or deal with highly symmetric configurations where the flow has a geometric (local) interpretation. In a generic case however, we expect to see many forms of non-localities. An initially local operator that is subject to non-local modular flow acquires contributions from spacelike separated regions with increasing modular time. Quantum entanglement of spacetime seems to play a crucial role here. Therefore it is of great interest to obtain further detailed understanding of explicit realizations of modular flow  for specific cases.

The example of the chiral fermion is rich enough for understanding  non-universal behaviour in detail,
 but still simple enough for explicit computations. Recently, this has lead to novel results on the modular Hamiltonian and relative entropy for disconnected regions \cite{Casini:2009vk,Klich:2015ina,Klich:2017qmt,Fries:2019ozf,Fries:2019acy,PhysRevD.100.025003,Hollands:2019hje}. In this paper we go beyond these studies and derive  results for the modular flow itself. We provide explicit formulae that may find direct applications in studies of fermionic entanglement.

Let us briefly state our main results. We consider density matrices reduced to an arbitrary set of disjoint intervals $V = \bigcup_n [a_n,b_n]$. Modular flow of a fermion operator localised at $y\in V$ is given by the convolution
\begin{align}\label{eq:psi_1}
\sigma_t\left( \psi^\dag(y) \right)=\int_V\dif{x}\psi^\dag(x) \Sigma_t(x,y) \, ,
\end{align}
where the kernel is a function of the correlator, 
\begin{align}\label{}
\Sigma_t=\left( \frac{1-G|_V}{G|_V} \right)^{\ii t}.
\end{align}
Here the reduced propagator $G|_V$ is understood as a linear operator acting on smooth functions via convolution. This approach avoids the computation of the modular Hamiltonian and instead directly yields the flow. We determine this kernel, and the explicit formulae for the modular flow are given in \eqref{sigmasol} (for the plane or cylinder (A)), \eqref{sigma_R} (cylinder (P)) and \eqref{psi_t_torus} (torus), which are illustrated in figures \ref{fig:two_ints}, \ref{fig:Ramond}, and \ref{fig:torus} respectively. Here, the notation P and A refers to periodic and antiperiodic boundary conditions for the fermions on the spatial circle, in other words the Ramond and Neveu-Schwarz sector. As a second novel result, we explicitly compute the modular two-point function, which is given by
\begin{align}\label{eq:Gmod_1}
G_{\text{mod}}(x,y;t)=\langle \psi(x)\, \sigma_t( \psi^\dag(y))\rangle\,.
\end{align}

The final results are, in the same order as above, \eqref{eq:g-mod12}, \eqref{eq:g-mod-cyl} and \eqref{eq:g-mod-torus}. A remarkable feature of this approach is that, although $\sigma_t(\psi^\dag)$ generically involves solving higher-degree polynomials or transcendental equations, the modular two-point function can be determined analitically by direct integration, without the need of solving such equations.

An important concept in our paper will be that of different degrees of \textit{locality} of the modular flow. This can be understood directly from \eqref{eq:psi_1}. We call a flow \textit{completely non-local} if the kernel $\Sigma_t(x,y)$ is a smooth function of $x,y$ supported on the entire region $V$, since it mixes operators along the entire region. If however $\Sigma_t(x,y)\sim \delta \left( f(x,y) \right)$ for some function $f$, the integral in \eqref{eq:psi_1} will localise to a  discrete set of isolated contributions, namely the zeroes of $f$. Generically, these solutions will be non trivial, in the sense that $x\neq y$ at $t=0$. We call these solutions \textit{bi-local}, since they couple pairs of distinct points. Finally, if there is a solution such that $x=y$ at $t=0$, we call it \textit{local}. We will use this terminology throughout the text. As we shall see, one of the essential features of our operator flow \eqref{eq:psi_1} is that it changes its locality properties depending on the temperature and the spin boundary conditions. In turn, this manifests itself in the structure of poles and cuts of the modular correlator \eqref{eq:Gmod_1}. 


The  paper is organized as follows. In section \ref{sec:modular-flows-overview} we specify the objects that we aim to compute: the modular flow of fermion operators and the modular two-point function. To this effect,  we first review some basic notions of Tomita-Takesaki theory, a mathematical framework relevant for local quantum field theories.  In section \ref{subsec:mod_flow_fermions} we introduce the particular physical system we focus on -- the free chiral fermion -- together with the necessary tools we will use to study modular flow. The main ingredient is holomorphic functional calculus, including the method of the resolvent. Section \ref{subsec:resolvent} includes a list of all known fermion resolvents. Section \ref{sec:op_flow} contains the first new results of this paper: We apply the above techniques to find the modular flow of the fermion operator for all cases considered.
Section \ref{sec:mod_two} presents our second important result, the modular two-point function. We verify that it obeys all required properties, such as analyticity and the KMS condition. We conclude by a summary and future directions in section \ref{sec:conclusions}.



\section{Modular flows: An overview}
  \label{sec:modular-flows-overview}
\subsection{A lightning overview of modular theory}
Before we specialize to the free fermion, let us first recall some basic notions of
(Tomita-Takesaki) modular theory. Consider a von Neumann algebra $\cR$ in normal form,
i.e. acting on a Hilbert space $\cH$ with a cyclic separating vector $\gns$. We
can then define the Tomita conjugation $S$ by
\begin{equation}
  S \cO\gns := \cO^\dagger\gns
\end{equation}
for all operators $\cO\in\cR$. Since $\gns$ is cyclic and separating, this defines $S$ on
a dense subspace of $\cH$. Furthermore, one can show that $S$ is closable, hence, admits a
unique polar decomposition
\begin{equation}
  S = J \Delta^{1/2}\,,
\end{equation}
where $J$ is antiunitary and $\Delta$ is positive. We denote $J$ and $\Delta$ as the
modular conjugation and modular operator, respectively.

The importance of these operators stems from Tomita's theorem, which states that
\begin{equation}
  J \cR J^\dagger = \cR' \qtext{and} \Delta^{\ii t} \cR \Delta^{-\ii t} = \cR\,,
\end{equation}
i.e. $J$ intertwines $\cR$ and its commutant $\cR'$ (the set of bounded operators in $\cH$ that commute with those in $\cR$) while the modular flow
\begin{equation}
  \sigma_t(\cO) := \Delta^{\ii t} \cO \Delta^{-\ii t}
\end{equation}
preserves $\cR$. Hence we see that modular flow is a
symmetry of $\cR$ arising just from the algebraic structure. The generator $K$ of this
symmetry, in the sense that
\begin{equation}
  \ee^{-\ii t K} := \Delta^{\ii t}\,,
\end{equation}
is referred to as the modular Hamiltonian. The reason for this is that, just like time
evolution, modular flow obeys a Kubo-Martin-Schwinger (KMS) condition: For
$\cO_1,\cO_2 \in \cR$, the modular correlation functions
\begin{equation}\label{eq:strips}
  \gnsev{\cO_1 \sigma_t(\cO_2)} \qtext{and} \gnsev{\sigma_t(\cO_2) \cO_1}
\end{equation}
admit analytic continuations to the strips $-1 \leq \Im(t) \leq 0$ and
$0 \leq \Im(t) \leq 1$, respectively, where they satisfy
\begin{equation}\label{eq:kms}
  \gnsev{\cO_1 \sigma_t(\cO_2)} = \gnsev{\sigma_{t+\ii}(\cO_2) \cO_1}\,.
\end{equation}

This is to be compared with the original KMS condition for a state
$\rho = \cZ^{-1} \ee^{-\beta H}$ of inverse temperature $\beta$: Denoting time evolution by
$\alpha_t(\cO) := \ee^{\ii t H} \cO \ee^{-\ii t H}$, we have
\begin{equation}
  \Tr[\rho \cO_1 \alpha_t(\cO_2)] = \cZ^{-1} \Tr[\ee^{-\ii(t-\ii\beta) H} \cO_1 \ee^{\ii t H} \cO_2] = \Tr[\rho \alpha_{t-\ii\beta}(\cO_2)\cO_1]\,.
\end{equation}
Therefore, $\sigma_t$ behaves like a time evolution with respect to the Hamiltonian $-K$
in a state of inverse temperature $-1$. In other words: If we only consider operators in
$\cR$, the vector $\gns$ behaves like a thermal state $\rho_\cR = \cZ^{-1} \ee^{-K}$. This
coincides with the definition of modular flow in terms of ``reduced density
matrices'' \cite{Witten:2018zxz}. As a word of caution, we would like to mention here that
reduced density matrices do not exist as operators in a genuine quantum field theory (QFT)
due to the universal divergence of vacuum
entanglement \cite{Haag:1992hx,Brattelli:1997fuh,Witten:2018zxz}. However, in the
following, we will use the formal analogy to finite dimensional systems (where reduced
density matrices do exist) to derive relevant formulae. While not entirely rigorous, this
method has proven to be very useful in the past and was confirmed in many
cases \cite{Araki:1971id,Hollands:2019hje}.

We now specialize to QFT in flat spacetime: In the Haag-Kastler approach to QFT, a von
Neumann algebra $\cR$ is associated to each (causally complete) region in space time,
typically denoted by the same symbol. This algebra can be thought of as consisting of the set of
(bounded) operators that have support in $\cR$. Since the associated modular flow
preserves this algebra and, hence, the associated region, it is tempting to ask, to which
extent it has a geometric or physical meaning. Remarkably, this question has an affimative
answer in the following scenario: If $\gns$ is the vacuum state and $\cR$ a Rindler wedge,
then $K$ is nothing but the (approriately scaled) generator of Lorentz boots that
preserves this wedge \cite{Bisognano:1975ih}. Furthermore, we can sometimes use additional
symmetries (e.g. conformal symmetry) to generalize this geometric action to other regions
such as a lightcone, double cones, or even to other states such as the thermal state \cite{Hislop1982zhn}.

Finally, let us elaborate further on the KMS condition in the context of fermionic
theories. As an example, we consider eq.~\eqref{eq:kms} for the case of two field operators
$\cO_1 = \psi(x)$ and $\cO_2 = \psi^\dagger(y)$. As stated, the two functions defined on \eqref{eq:strips} are analytic when restricted to the lower and upper unit strips respectively. Therefore, it is natural to define the \textit{modular} two-point function \cite{Haag:1967zfg,Haag:1992hx,Hollands:2019hje}
\begin{equation}\label{eq:master}
  G_{\text{mod}}(x,y;t) :=
  \begin{cases}
   -\gnsev{\sigma_t(\psi^\dagger(y))\psi(x)} &\text{for } 0 < \Im(t) < 1\\
    +\gnsev{\psi(x) \sigma_t(\psi^\dagger(y))} &\text{for } -1 < \Im(t) < 0.
  \end{cases}
\end{equation}
This function, defined via a \textit{different} function in each strip, satisfies the
following antiperiodicity due to the KMS condition \eqref{eq:kms}
\begin{equation}
  \label{eq:kms2}
  G_{\text{mod}}(x,y;t) = - G_{\text{mod}}(x,y;t+i)\ \ ,\ \ -1 < \Im(t) < 0
\end{equation}
allowing to continue $\Gmod$ to arbitrary non-integer imaginary parts of
$t$.

The reason to define \eqref{eq:master} is because of its relation to the
anticommutator. While $G_{\text{mod}}$ is analytic in the lower and upper strips by
construction, the interesting question regards its regularity properties of along
$\Im (t)=0$ (and, hence, at all integer imaginary parts due to antiperiodicity). Now it is
easy to see that the variation of $\Gmod$ along the real axis is given by the
anticommutator between a field and the modular evolved field:
\begin{equation}\label{eq:dGmod}
  G_{\text{mod}}(x,y;t-\ii 0^+) - G_{\text{mod}}(x,y;t+\ii 0^+) = \gnsev{\{\psi(x), \sigma_t(\psi^\dagger(y))\}}\ \ ,\ \ t\in\mathbb R
\end{equation}

This condition becomes particularly useful in cases where the modular evolution is given
by a smearing of the field (this will be precisely the case for gaussian free fermion
states)
\begin{equation}
  \sigma_t\big(\psi^\dagger(y)\big) = \int_V \dif[d]x \psi^\dagger(x) \Sigma_t(x,y).
\end{equation}
with which \eqref{eq:dGmod} becomes
\begin{equation}
  G_{\text{mod}}(x,y;t-\ii 0^+) - G_{\text{mod}}(x,y;t+\ii 0^+) = \Sigma_t(x,y)
\end{equation}
by the canonical anticommutation relation
$\{\psi(x),\psi^\dagger(y)\}=\delta(x-y)$. Equivalently, we can use the antiperiodicity
\eqref{eq:kms2} to rewrite this purely in terms of the function on the lower strip,
\begin{equation}
  \label{eq:kms3}
  G_{\text{mod}}(x,y;t-\ii 0^+) + G_{\text{mod}}(x,y;t-\ii+\ii 0^+) = \Sigma_t(x,y).
\end{equation}

This relation is important because it relates the analytic structure of the modular
correlator to the locality properties of the modular flow, via the Kernel $\Sigma_t$ of
the operator flow. If the flow under consideration is local, i.e., if
$\Sigma_t(x,y) \propto \delta(x-y)$, the right hand side vanishes almost everywhere and we
obtain antiperiodicty of the two-point function in imaginary time. On the other hand, this
means that every failure of such regularity (e.g. a branch cut of $G_{\text{mod}}$) is a
clear sign of non-locality in the modular flow.

In section \ref{sec:mod_two} we will explicitly determine $\Gmod(t)$ for the
two-dimensional chiral fermion. We will confirm that for the cases where modular flow is
local or bi-local (such as a single interval on the plane, antiperiodic cylinder, or
torus), the modular correlator restricted to the lower strip yields a function that is
analytic everywhere away from isolated simple poles. The salient example where analyticity
fails is a single interval on the periodic vacuum on the cylinder, where the flow is
completely non-local and $\Gmod$ has branch cuts.

\subsection{Modular flows for free Fermions}
\label{subsec:mod_flow_fermions}
For the rest of the paper, we restrict to Gaussian states in a Fermionic theory. Also, we are
working on a Cauchy slice and consider only subregions
$V$ of that slice. In this case, we can formally decompose the modular Hamiltonian as
\begin{equation}
  K = \int_V \dif[d]x \int_V \dif[d]y \psi^\dagger(x)k(x,y)\psi(y) + \int_{V^c} \dif[d]x \int_{V^c} \dif[d]y \psi^\dagger(x)k^c(x,y)\psi(y)\,,
\end{equation}
where $V^c$ is the complement of $V$ in the Cauchy slice. The absence of mixing terms
between $V$ and $V^c$ reflects the fact that modular flow preserves $\cR$ and
$\cR'$, which are associated to $V$ and $V^c$ respectively. Mathematically, the above Kernels $k, k^c$ have to be understood in the
distributional sense, i.e. they are only defined as integrated against suitably smooth
test functions. For the remained or the text we restrict to $k$, as the calculation of $k^c$ is completely analogous.

To derive an explicit formula for $k$, we require that the ``reduced density
matrix'' $\cZ^{-1} \ee^{-K}$ reproduces the correct expectation values for operators with
support in $V$. By Wick's theorem it is enough to reproduce the propagator
\begin{equation}
  G(x,y) := \gnsev{\psi(x)\psi^\dagger(y)}
\end{equation}
in the subregion $x,y\in V$, which allows to derive the relation \cite{Araki:1971id,Peschel:2003xzh}
\begin{equation}
  \ee^{-k} = \frac{1-G|_V}{G|_V}\,,
\end{equation}
where $G|_V$ is the restriction of $G$ to $V$. In a similar manner, we arrive at
\begin{equation}\label{tomitapsi}
  \sigma_t\big(\psi^\dagger(y)\big) = \int_V \dif[d]x \psi^\dagger(x) \Sigma_t(x,y) \qtext{with} \Sigma_t = \bigg[\frac{1-G|_V}{G|_V}\bigg]^{\ii t}
\end{equation}
and thus (for $-1 < \Im(t) \leq 0$)
\begin{equation}
  \label{eq:mod2pt}
  G_\mathrm{mod}(x,y;t) = \gnsev{\psi(x)\sigma_t\big(\psi^\dagger(y)\big)} = \bigg(G|_V\bigg[\frac{1-G|_V}{G|_V}\bigg]^{\ii t}\bigg)(x,y)\,.
\end{equation}
Here and below, we use the compact notation $\Sigma_t$ omitting its space-time dependence,
but it should always be kept in mind that it is a linear operator acting on
functions. Similarly for other operators such as the resolvent.

\begin{figure}[h]
  \def\svgwidth{.7\linewidth}
  \centering{
  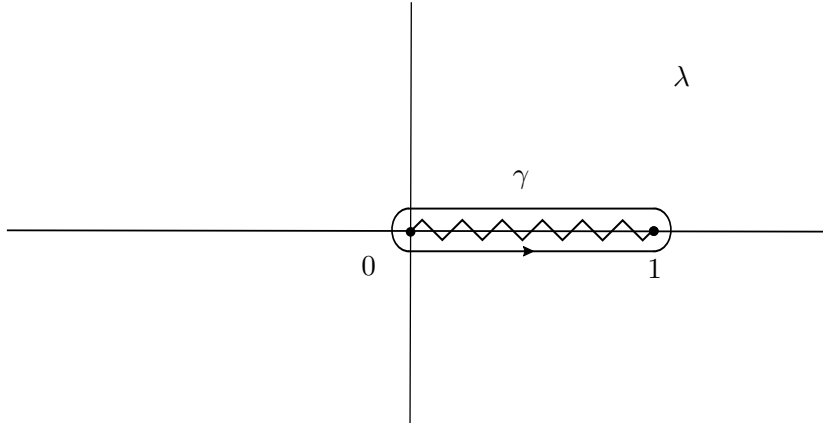
  \caption{Contour used in \eqref{eq:cauchy}. The resolvent $(\lambda-G)^{-1}$ must
    possess a cut along $[0,1]$, the spectrum of $G$.}
  \label{fig:cont_1}}
\end{figure}

The problem is thus reduced to computing functions of the (restricted) propagator
$G|_V$. Since this is a bounded operator -- its spectrum is contained in the interval
$[0,1]$ -- we can use functional calculus to write
\begin{equation}
  \label{eq:cauchy}
  f(G|_V) = \frac 1{2\pi\ii} \oint_{\gamma} \dif \lambda f(\lambda) \frac 1{\lambda-G|_V}\,,
\end{equation}
where $1/(\lambda-G|_V)$ is the resolvent of $G|_V$ and $\gamma$ denotes that the integral
is to be done counter-clockwise along a contour that tightly wraps around the spectrum
$[0,1]$, as shown in figure~\ref{fig:cont_1}. Eq.~\eqref{eq:cauchy} can easily be seen to
be correct in an eigendecomposition of $G|_V$ and implies that the resolvent
$1/(\lambda-G|_V)$, as a function of $\lambda$, is analytic in a neighbourhood of $[0,1]$,
but not along the interval itself: If the spectrum of $G|_V$ is discrete, we expect a
simple pole whenever $\lambda$ approaches an eigenvalue. For continuous portions of the
spectrum, this culminates in a branch cut. In any closed form expression of the resolvent,
such a branch cut will only be visible as a pair of branch points and one might be tempted
to assume that the precise location of the cut is indeterminate. However, as just
discussed, one should always keep in mind that the branch cut will always be situated
alongside $[0,1]$.

Let us compare eq.~\eqref{eq:cauchy} with the general definition of a function of the kernel $G|_V$, given by
\begin{equation}
  \label{spectral}
  f(G|_V) = \int_V \dif{E_\lambda} f(\lambda) = \int_V \dif\lambda \od{E_\lambda}{\lambda} f(\lambda)\,,
\end{equation}
where $E_\lambda$ is the spectral measure of $G|_V$. Assuming the contributions to
eq.~\eqref{eq:cauchy} at $0$ and $1$ vanish, we obtain
\begin{equation}
  \od{E_\lambda}{\lambda} =
  \frac 1{2\pi\ii} \bigg[\frac 1{\lambda-G|_V-\ii 0^+} - \frac 1{\lambda-G|_V+\ii 0^+}\bigg],
\end{equation}
which characterizes the spectral measure completely. Note that the requirement of
vanishing contributions at $0$ and $1$ also imposes a regularity constraint on the
function $f$. For the computations in section~\ref{sec:op_flow}, it will turn our that
this contraint is violated and we will have to work with eq.~\eqref{eq:cauchy} directly.
 
To proceed any further, it is necessary to find the resolvent for the state and region under
consideration. To this end, we can make the ansatz
\begin{equation}
  \label{eq:res}
  \frac 1{\lambda -G|_V} = \frac 1\lambda + \frac{F_\lambda}{\lambda^2}\,,
\end{equation}
which turns the functional equation $(\lambda-G_V) \times 1/(\lambda-G_V) = 1$ into the
integral equation
\begin{equation}
  \label{eq:integral-equation}
  -G(x,y) + F(x,y) - \frac 1\lambda \int_V\dif[d]z G(x,z)F_\lambda(z,y) = 0\,, \quad x,y\in V\,.
\end{equation}
Notice that while this equation is valid for fermions in arbitrary dimensions, the
solutions are only known in two dimensions, which is the focus of the next subsection.

\subsection{Resolvent for the chiral Fermion}
  \label{subsec:resolvent}
Due to recent developments \cite{Fries:2019ozf}, \eqref{eq:integral-equation} is well
understood in the special case where we are dealing with a chiral fermion in one dimension
and $V = \bigcup_n [a_n,b_n]$ is a finite union of disjoint intervals. This is because
there, the propagator is a Cauchy kernel, hence the integral equation can be reduced to yet another complex analysis problem, where the resolvent has a branch cut along $V$. We omit details here and just state the results in the
following cases, classified by the domain/periodicities of $G$:
\begin{itemize}
\item No periodicity (the entire complex plane)---the corresponding propagator is given by
  \begin{equation}
    \label{eq:g1}
    G(x,y) = \frac 1{2\pi\ii}\frac 1{x-y-\ii 0^+}\,.
  \end{equation}
  The solution is
  \begin{equation}
    \label{eq:f1}
    F_\lambda(x,y) = - \frac\lambda{1-\lambda} G(x,y) \bigg[- \frac{1-\lambda}\lambda\bigg]^{\ii [Z(x)-Z(y)]}\,,
  \end{equation}
  with
  \begin{equation}
    \label{eq:z1}
    Z(x) = \frac 1{2\pi} \log \bigg[-\prod_n \frac{a_n-x}{b_n-x}\bigg]\,.
  \end{equation}
\item One periodicity, taken to be $1$ without loss of generality (the complex cylinder)
  -- the propagators are
  \begin{align}
    G(x,y) &= \frac 1{2\ii} \csc \pi(x-y-\ii 0^+) \label{eq:g2}\,, \\
    G(x,y) &= \frac 1{2\ii} \cot \pi(x-y-\ii 0^+) \label{eq:g3}\,,
  \end{align}
  depending on the choice of antiperiodic or periodic boundary conditions,
  respectively. The corresponding solutions are
  \begin{align}
    F_\lambda(x,y) &= - \frac\lambda{1-\lambda} G(x,y) \bigg[- \frac{1-\lambda}\lambda\bigg]^{\ii [Z(x)-Z(y)]}\,, \label{eq:f2} \\
    F_\lambda(x,y) &= - \frac\lambda{1-\lambda} \bigg[G(x,y) + \frac 12 \frac{[-(1-\lambda)/\lambda]^L-1}{[-(1-\lambda)/\lambda]^L+1}\bigg]
                     \bigg[- \frac{1-\lambda}\lambda\bigg]^{\ii [Z(x)-Z(y)]}\,, \label{eq:f3}
  \end{align}
  with the total length $L = \sum_n (b_n-a_n)$ of $V$ and
  \begin{equation}
    \label{eq:z23}
    Z(x) = \frac 1{2\pi} \log \bigg[-\prod_n \frac{\sin \pi(a_n-x)}{\sin \pi(b_n-x)}\bigg]\,.
  \end{equation}
\item Two periodicities $1, \tau$ (the complex torus) -- here, the propagators are
  \begin{equation}
    \label{eq:g4}
    G^{(\nu)}(x,y;\tau) = \frac{\eta^3(\tau)}{\ii\vartheta_1(x-y-\ii 0^+|\tau)} \frac{\vartheta_\nu(x-y|\tau)}{\vartheta_\nu(0|\tau)}
  \end{equation}
  with $\nu=2,3$ denoting the periodic-antiperiodic (PA) and antiperiodic-antiperiodic
  (AA) boundary conditions, respectively. The conventions for Jacobi theta and Dedekind eta functions are the same as in \cite{DiFrancesco:1997nk}. The solutions of \eqref{eq:integral-equation}
  are now
  \begin{align}
    F^{(\nu)}_\lambda(x,y) &= - \frac\lambda{1-\lambda} G^{(\nu)}(x,y;\tau,Lh) \bigg[- \frac{1-\lambda}\lambda\bigg]^{\ii [Z(x)-Z(y)]}\,, \label{eq:f4} \\
  \end{align}
  where $h$ is defined by $\ee^{2\pi h} := -\frac{1-\lambda}\lambda$ and
  \begin{align}    
    G^{(\nu)}(x,y;\tau,\mu) &=  \frac{\eta^3(\tau)}{\ii\vartheta_1(x-y-\ii 0^+|\tau)} \frac{\vartheta_\nu(x-y-\ii\mu|\tau)}{\vartheta_\nu(-\ii\mu|\tau)}\,, \\
    Z(x) &= \frac 1{2\pi} \log \bigg[-\prod_n \frac{\vartheta_1(a_n-x|\tau)}{\vartheta_1(b_n-x|\tau)}\bigg] \label{eq:z4}\,.
  \end{align}
  Note that $G^{(\nu)}(x,y;\tau,Lh)$ is the propagator of a state with chemical potential $Lh$, i.e. we have the series representations
  \begin{align}    
    G^{(2)}(x,y;\tau,Lh) &= \sum_{k\in\ZZ}' \frac{\ee^{-2\pi\ii k(x-y-\ii 0^+)}}{1+\ee^{2\pi (\ii k \tau-Lh)}} = \sum_{k\in\ZZ}' \frac{\ee^{-2\pi\ii k(x-y-\ii 0^+)}}{1+[-(1-\lambda)/\lambda]^{-L}\ee^{2\pi \ii k \tau}}\,, \label{eq:gg2}\\
    G^{(3)}(x,y;\tau,Lh) &= \sum_{k\in\ZZ+1/2}' \frac{\ee^{-2\pi\ii k(x-y-\ii 0^+)}}{1+[-(1-\lambda)/\lambda]^{-L}\ee^{2\pi \ii k \tau}}\,, \label{eq:gg3}
  \end{align}
  where the symbol $\sum'$ denotes that the sums have to be ordered symmetrically to ensure convergence.
\end{itemize}


\section{Modular flow of operators}
\label{sec:op_flow}

In this section we will compute explicitly the modular flow of the fundamental field, $\sigma_t(\psi^\dag)$ from \eqref{tomitapsi}. This is a basic building block that allows to compute the flow of composite operators.
As explained in section \ref{subsec:mod_flow_fermions}, the task reduces to determining the kernel 
\begin{align}\label{eq:Sigmat0}
\Sigma_t=\left( \frac{1-G|_V}{G|_V} \right)^{\ii t}\,.
\end{align}
Using the Cauchy formula \eqref{eq:cauchy}, with $f(\lambda)=\left( \frac{1-\lambda}{\lambda} \right)^{\ii t}$ and decomposing the resolvent as in \eqref{eq:res} we have
\begin{equation}\label{Sigmat1}
  \Sigma_t = \frac 1{2\pi\ii} \oint_\gamma  \dif\lambda \left( \frac{1-\lambda}{\lambda} \right)^{\ii t} \left[ \frac{1}{\lambda}+\frac{F_\lambda}{\lambda^2} \right]\,.
\end{equation}

As it stands, this integral is not completely well defined, as the integrand is both divergent and highly oscillating around the branch points. However, this should come as no surprise: since we know that $\Sigma_t(x,y)$ represents a distribution, we expect the appearance of Dirac delta distributions. Thus the strategy is to regularise the integral, evaluate it, and finally remove the regulator  and identify the remaining distributions. 

Here the analytic structure of the integrand is crucial. In addition to the cut associated to the resolvent -- the last factor in \eqref{Sigmat1} -- $f(\lambda)$ has introduced another cut, branched over the same endpoints. The latter cut can be freely chosen as long as it does not overlap with the former. For simplicity, we choose it to run along the real complement, $\RR \setminus  [0,1]$ -- see fig. \ref{fig:cont_2}. 

In appendix \ref{appendixA} we provide a rigorous treatment of this integral and evaluate it by residues. Here instead, we proceed with a more straightforward but nevertheless equivalent approach. As explained in appendix \ref{appendixA}, a standard regularisation consists of avoiding the poles at $\lambda=0$ and $\lambda=1$ by shifting them slightly into the complex plane. As a consequence, the integral with the first term in the square brackets, proportional to $1/\lambda$, vanishes. This can be seen as follows. This term has a branch cut along $(-\infty,0)\,\cup\,(1,\infty)$ and a pole at $\lambda=0$, but everywhere else is holomorphic. In particular, the contribution around $\lambda=1$ vanishes due to the KMS requirement. Since both boundary terms vanish, the integral vanishes. For more details, please refer to appendix \ref{appendixA}. Thus, we are left with
\begin{equation}\label{Sigmat2}
 \Sigma_t = \frac 1{2\pi\ii} \oint_\gamma \dif\lambda \left( \frac{1-\lambda}{\lambda} \right)^{\ii t} \frac{F_\lambda}{\lambda^2}\,.
\end{equation}
Finally, $F_\lambda$ takes a different form depending on the topology and boundary conditions chosen. We consider them case by case. 

\subsection{Plane}
\label{subsec:vacuumflow}

We start with the simplest case. For the vacuum state on the plane \eqref{eq:f1}, $F$ takes the form
\begin{align}\label{F_lam}
F_\lambda(x,y)=-\frac{\lambda}{1-\lambda} G(x,y) \left( - \frac{1-\lambda}{\lambda} \right)^{\ii \tilde t}\,,
\end{align}
where we have introduced the shorthand notation
\begin{align}\label{tildet}
\tilde t(x,y)=Z(x)-Z(y)
\end{align}
and throughout the text we use $\tilde t=\tilde t(x,y)$ and omit the spacetime dependence, which should nevertheless be kept in mind. The notation will become clear shortly, as we will see that $\tilde t$ plays a role closely analogous to modular time $t$. 

\begin{figure}[h]
	\def\svgwidth{.7\linewidth}
	\centering{
		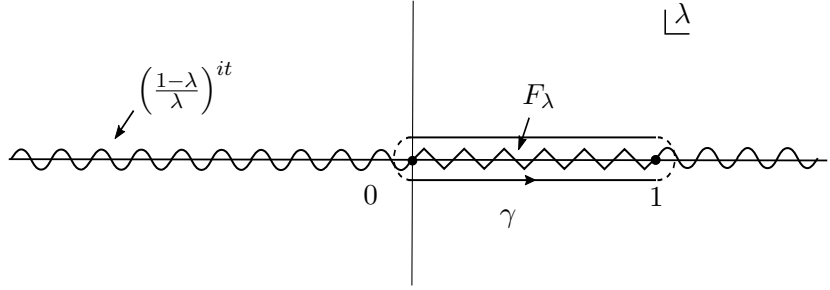
		\caption{ Contour for the integral \eqref{Sxy}, with the associated branch cuts indicated. As explained in the text, the regularisation yields the integrand bounded around the origin, and thus the contribution the along the dashed semicircles vanish. }
                \label{fig:cont_2} 	}
\end{figure}

Also, it is important to realise that here, the propagator $G(x,y)$ has no dependence on $\lambda$ and therefore can be pulled out of the integral (notice however that this will not hold for the cylinder (R) or torus). Introduced back in \eqref{Sigmat2}, one obtains
\begin{align}\label{Sigma_t}
\Sigma_t(x,y)&=G(x,y)S(x,y)\,,
\end{align}
where we have defined the integral (see fig. \ref{fig:cont_2})
\begin{align}\label{Sxy}
S(x,y)=-\frac{1}{2\pi\ii}\oint_{\gamma} \frac{\dif\lambda}{\lambda(1-\lambda)} \left( \frac{1-\lambda}{\lambda} \right)^{\ii t} \left( - \frac{1-\lambda}{\lambda} \right)^{\ii \tilde t}\,.
\end{align}

\begin{figure}[h]
	\def\svgwidth{.7\linewidth}
	\centering{
		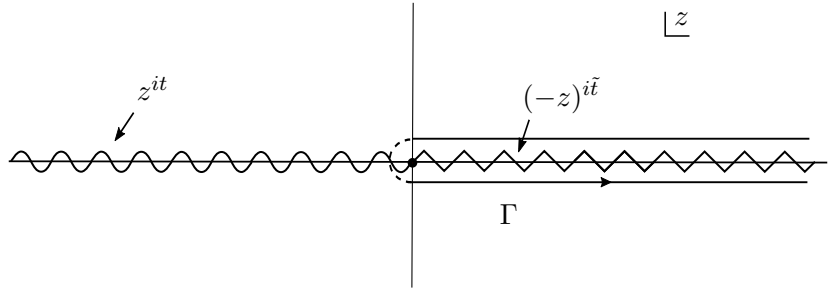
		\caption{ Contour for \eqref{Sxy2}, obtained from the previous figure by $z=(1-\lambda)/\lambda$. Again, the contribution of the dashed segment vanishes when properly regularised. }
                \label{fig:cont_3} 	}
\end{figure}

In order to exploit the symmetry and simplify the integral, it is useful to change variable to $z=(1-\lambda)/\lambda$, which maps the cut along $[0,1]$ to $\RR^{+}$, and $\RR\setminus [0,1] $ to  $\RR^{-}$, while the image of the contour wraps positively around $\RR^{+}$, see fig. \ref{fig:cont_3}.  To make the branch cuts explicit, we use $-\pi<\arg z<\pi$ for the cut along $\mathbb R^-$ and $0<\arg z<2\pi$ for the one along $\mathbb R^+$. The integral \eqref{Sxy} now reads
\begin{align}\label{Sxy2}
S(x,y)&=\frac{1}{2\pi \ii}\oint_{\Gamma} \frac{\dif{z}}{z} z^{\ii t} (-z)^{\ii\tilde t}\,.
\end{align}
The discontinuity along $\mathbb R^+$ implies that just above and below the cut we have
\begin{align}\label{}
\left( -(z\pm \ii 0^+) \right)^{\ii\tilde t}=\ee^{\pm \pi \tilde t} z^{\ii\tilde t}\,.
\end{align}
As mentioned above and discussed in detail in the appendix, the regularisation makes the integrand bounded in a neighbourhood of the origin $z=0$, so we can neglect the contribution of the boundary point. Putting everything together, \eqref{Sxy2} can be formally represented as 
\begin{align}\label{}
S(x,y)&=-\frac{1}{2\pi \ii}\left( \ee^{\pi \tilde t}-\ee^{-\pi \tilde t} \right)\int_{\mathbb R^+} \dif{z}\, z^{\ii (t+\tilde t)-1}\,.
\end{align}

Now we proceed to the regularisation. This integral receives divergent contributions from both $z\to 0$ and $z\to \infty$. Both can be made finite by restricting the integration to $z\in(\ee^{-2\pi m},\ee^{2\pi m})$ where we eventually take $m\to \infty$. The result is
\begin{align}\label{eq:Ssin}
S(x,y)=2\ii\sinh \left( \pi \tilde t \right)  \frac{\sin 2\pi m(t+\tilde t)}{\pi (t+\tilde t)}\,.
\end{align}
Again, we remind the reader that a rigorous derivation of this via residue analysis is presented in appendix \ref{appendixA}. This expression is actually familiar. If $t+\tilde t(x,y)\neq 0$, the fraction is bounded but wildly oscillating, and therefore vanishes when integrated agains regular test functions. In the vicinity of $t+\tilde t(x,y)=0$ instead, the fraction diverges as $m\to\infty$. This is the standard Dirichlet kernel representation of the Dirac distribution, so
\begin{align}\label{Sxy3}
S(x,y)=-2\ii \sinh (\pi t) \delta \left( t+\tilde t \right)\,.
\end{align}

As stated above, it is clear that $\tilde t$ is indeed playing a role somewhat analogous to modular time itself. Putting back all together into \eqref{Sigma_t} and replacing \eqref{tildet}, we learn that the kernel associated to the action of modular flow is
\begin{align}\label{Sigmasol}
\Sigma_t(x,y)=-2\ii \sinh(\pi t) G(x,y)\delta(t+Z(x)-Z(y))\,,
\end{align}
whose support is given by the solutions of
\begin{align}\label{tdZ}
t+\tilde t =t+Z(x)-Z(y)=0\,.
\end{align}
This equation, and its solutions, play a fundamental role in our analysis. It will determine which points are non-locally coupled via modular flow, as well as the magnitude of their coefficients. For a fixed $y$, we shall call $x_\ell=x_\ell(y)$ the solutions for $x$, where the discrete index $\ell$ labels the different intervals in $V$. The most important property of the function $Z$ is that it increases monotonically from $-\infty$ at each left endpoint $a_j$ to $+\infty$ at the right endpoints $b_j$. This guarantees that there will exist one solution to this equation per interval. 

The action of modular flow finally reads
\begin{align}\label{sigmasol}
\sigma_t\left( \psi^\dag(y) \right)&=-2\ii \sinh(\pi t) \sum_\ell \frac{G(x_\ell,y)}{Z'(x_\ell)} \psi^\dag(x_\ell)\,,
\end{align}
where as before $x_\ell$ are the solutions of \eqref{tdZ}. We omit the absolute value in the denominator since $Z(x)$ is monotonically increasing in each interval. In order to gain some intuition, we illustrate these results below with some simple examples. 

To conclude here, let us note what happens in the limit $t\to 0$. At zero time, the kernel $\Sigma_t$ in \eqref{eq:Sigmat0} must reduce to the identity, localized at $x=y$. Now the prefactor in \eqref{sigmasol} vanishes linearly with time. Since the propagator has a (unique) simple pole at coincident points, all but the `local' solution, which obeys $x_\ell\to y$ at $t\to 0$, vanish in \eqref{sigmasol}.

\paragraph{Rindler space.}  This is the best known explicit case, which obeys a universal formula for the vacuum of any QFT on the Rindler wedge \cite{Bisognano:1975ih}. Physically this corresponds to the standard Unruh effect, where modular evolution is nothing but translations along the worldline of observers with constant acceleration. Here the entangling region is $V=\RR^+$, which can be obtained by taking the limit $b\gg a$ of the single interval on the plane \eqref{eq:z1}, yielding
\begin{align}\label{}
Z(x)-Z(y)=\frac{1}{2\pi} \log \frac{x}{y}\,.
\end{align}
The unique solution to \eqref{tdZ} is $x_1=\ee^{-2\pi t}y$, which inserted back into \eqref{sigmasol} leads to the geometric flow
\begin{align}\label{}
\sigma_t\left( \psi^\dag(y) \right)=\ee^{-\pi t} \psi^\dag\left( \ee^{-2\pi t}y \right)\,,
\end{align}
the prefactor being due to the transformation law of a spin $1/2$ field under Lorentz boosts. 

\paragraph{Multiple intervals on the plane.} Here the entagling region is an arbitrary set of disjoint intervals $V=\cup_{i=1}^n (a_i,b_i)$. This was the case solved in the seminal work \cite{Casini:2009vk}. However it is important to note how their strategy differs from ours. In \cite{Casini:2009vk} the authors first derived the modular Hamiltonian $K_V=-\log \rho_V$ and next used the associated Heisenberg equation $\partial_t \psi=\ii[\psi,K_V]$. This yields a set of coupled differential equations relating the different $\psi(x_\ell(t))$. On the other hand, we computed modular flow \textit{directly} in terms of the resolvent. This avoids using the modular Hamiltonian itself, and the need for the differential equation, and hands at once the solution.

\begin{figure}[h]
 \centering
  \includegraphics[width=0.6\textwidth]{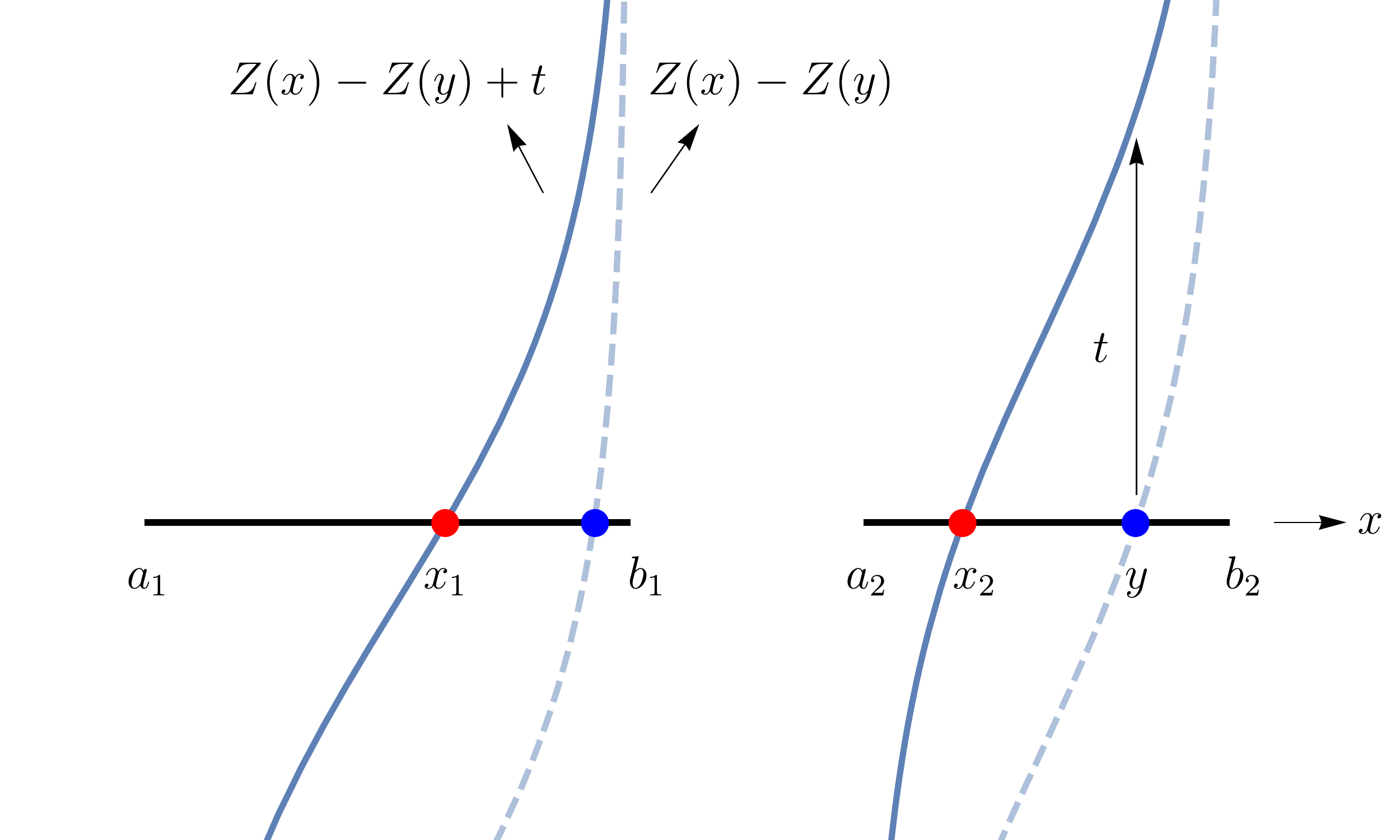}
 \caption{ An illustration of the modular flow $\sigma_t\left( \psi^\dag(y) \right)$ in \eqref{sigmasol} for the union of two intervals $(a_1,b_1)\cup(a_2,b_2)$ (black) on the plane. We plot the function $Z(x)-Z(y)+t$ for fixed $y$ as a function of $x$, for both $t=0$ (dashed curve) and $t>0$ (solid curve). The zeroes of the function correspond to the dots (blue for $t=0$, red for $t>0$). The flow couples the solutions of \eqref{tdZ}, $Z(x)-Z(y)+t=0$, which correspond to the zeroes of the function. Since $Z(x)$ increases monotonically from $-\infty$ to $\infty$ within each interval, there exists exactly one solution per interval. As modular time evolves, the curve is shifted linearly in time. Consequently the zeroes move left along modular flow. The solution contained in the same interval as $y$ is the `local' solution, while the other corresponds to the `non-local' one.  }
 \label{fig:two_ints}
 \end{figure}

For completeness, and in order to compare to the new results, we illustrate this case for two intervals $(a_1,b_1)\cup (a_2,b_2)$. Again, the solutions to \eqref{tdZ} are essential. In this case, 
\begin{align}\label{}
Z(x)=\frac{1}{2\pi} \log - \frac{(x-a_1)(x-a_2)}{(x-b_1)(x-b_2)}
\end{align}
and therefore \eqref{tdZ} leads to a second degree equation, which can be readily solved but the expression is rather cumbersome. We plot these solutions in Fig.\ref{fig:two_ints}. This was the first known case of a bi-local or \textit{quasi}-local modular flow that could be solved analytically \cite{Casini:2009vk}. The most important feature is that the flow contains involves two kinds of terms. The local solution lives in the same interval as $y$, and is continuously connected to it at $t=0$. But there are also bi-local terms, one per interval. Also notice that due to chirality, the solutions move towards the left as modular time $t$ evolves, converging to the left endpoints asymptotically, and similarly go to the right endpoints as $t\to -\infty$. 

This simple example illustrates an important point. In general, the configuration on the plane with $n$ intervals involves polynomials of degree $n$. Thus, solving explicitly for the modular flow of the fermion operator becomes quickly hopeless as we increase $n$. This becomes even more involved for other states like the cylinder or the torus. Now, since these terms will show up in the modular flow of composite operators, it would seem implausible to find any closed analytic results for those cases. However, as we will show in section \ref{sec:mod_two} with the modular two-point function, there exists a remarkable way to circumvent the need to solve \eqref{tdZ}. 

\subsection{Cylinder}
\label{subsec:vacuumP}

The vacuum state for the fermion on the cylinder possesses two spin sectors, the antiperiodic (A or Neveu-Schwarz) and the periodic (P or Ramond), depending on the boundary conditions we choose for the fermion along the circle. The modular flow for the antiperiodic sector (for any number of intervals) is identical to that of the plane, provided we use the appropriate propagator~\eqref{eq:g2} and $Z(x)$ given in \eqref{eq:z23}, and therefore we will not elaborate on it. However, this behaviour changes dramatically when we consider the periodic sector. 

The periodic sector on the cylinder provides an example of how the present method allows to go beyond previous results in the literature. This case must be considered separately, since the resolvent~\eqref{eq:f3} contains an extra term, due to the presence of a zero mode. The first term of \eqref{eq:f3} is of identical form as the one derived in the previous section, namely \eqref{Sigma_t}, where again $S(x,y)$ is given by \eqref{Sxy3} and the corresponding correlator on the cylinder (P), \eqref{eq:g3}. This allows us to decompose the modular evolution as 
\begin{align}\label{SigmaR}
\Sigma_t(x,y)=G(x,y) S(x,y) + \delta \Sigma_t\,,
\end{align}
where the extra term, associated to the Ramond sector, can be written in the variable $z=(1-\lambda)/\lambda$ as
\begin{align}\label{}
\delta \Sigma_t
&=\frac{1}{4\pi \ii}\oint_\Gamma \frac{\dif{z}}{z} z^{\ii t} \left( -z \right)^{\ii\tilde t} \frac{(-z)^L-1}{(-z)^L+1}
\end{align}
where again we defined $\tilde t=Z(x)-Z(y)$.

Using again \eqref{Sxy2}, this can be brought into the more convenient form
\begin{align}\label{}
\delta \Sigma_t&=\frac{1}{2}S(x,y) + \frac{1}{2\pi i}\oint_\Gamma \frac{\dif{z}}{z} z^{\ii t} \left( -z \right)^{\ii\tilde t} \frac{1}{(-z)^L+1}\,.
\end{align}
Although the last factor in the integral does not possess a multiplicative branch cut, one can bring it into such a form using the identity
\begin{equation}
  \label{eq:identity}
  \frac{1}{1+y}=\frac{\ii}{2}\int_{-\infty}^{\infty}\dif{s}\frac{y^{\ii s}}{\sinh\left(\pi s+\ii 0^+\right)}\quad\mathrm{for}\quad y\in\CC\setminus\RR^{-}\,,
\end{equation}
where we identify $y=(-z)^L$ for $L\in (0,1)$. Then, the same steps leading to \eqref{Sxy3} yield 
\begin{align}\label{}
\delta \Sigma_t
&=\frac{1}{2}S(x,y) + \frac{\sinh(\pi t)}{L\sinh\left( \frac{\pi (t+\tilde t)}{L} \right)}\,.
\end{align}

As we discuss below, this result is quite remarkable. While the first term -- given by \eqref{Sigmasol} -- produces a local flow, the second one does not and leads to complete non-locality. Indeed, as explained in \ref{sec:modular-flows-overview}, any contribution to the kernel $\Sigma_t(x,y)$ that is \textit{not} localised (in the sense of being proportional to $\delta(x-y)$) will produce a discontinuity in the modular two-point function, in accordance with the KMS condition \eqref{eq:kms2}. 

Finally, inserting back into \eqref{SigmaR}, \eqref{sigmasol} and \eqref{tomitapsi}, we find
\begin{align}\label{sigma_R}
\sigma_t\left( \psi^\dag(y) \right)=-2\ii \sinh(\pi t) \sum_\ell \frac{G(x_\ell,y)+1/2}{Z'(x_\ell)}\, \psi^\dag(x_\ell) + \frac{\sinh (\pi t)}{L} \int_V \dif{x} \frac{\psi^\dag(x)}{\sinh \frac{\pi}{L}(t+\tilde t)}\,.
\end{align}
The first term is very similar to the flow on the plane or cylinder (NS) of the previous section, but with the correlator shifted by a constant of $1/2$. This factor is not very significant as it can be reabsorbed into the integral; it will show up again in section \ref{subsec:mod2pt-cylinder}.  Again, this flow always involves a local (geometric) term, and additional bi-local (quasi-geometric) couplings whenever $V$ contains more than one interval. In particular, the solution for the $x_\ell$ is identical to that on the antiperiodic sector, since this depends only on the function $Z(x)$, which does not depend on the periodicity. Notice that, as mentioned above, at $t=0$ all terms vanish, except the one involving the propagator. 

\begin{figure}[h]
 \centering
  \includegraphics[width=0.6\textwidth]{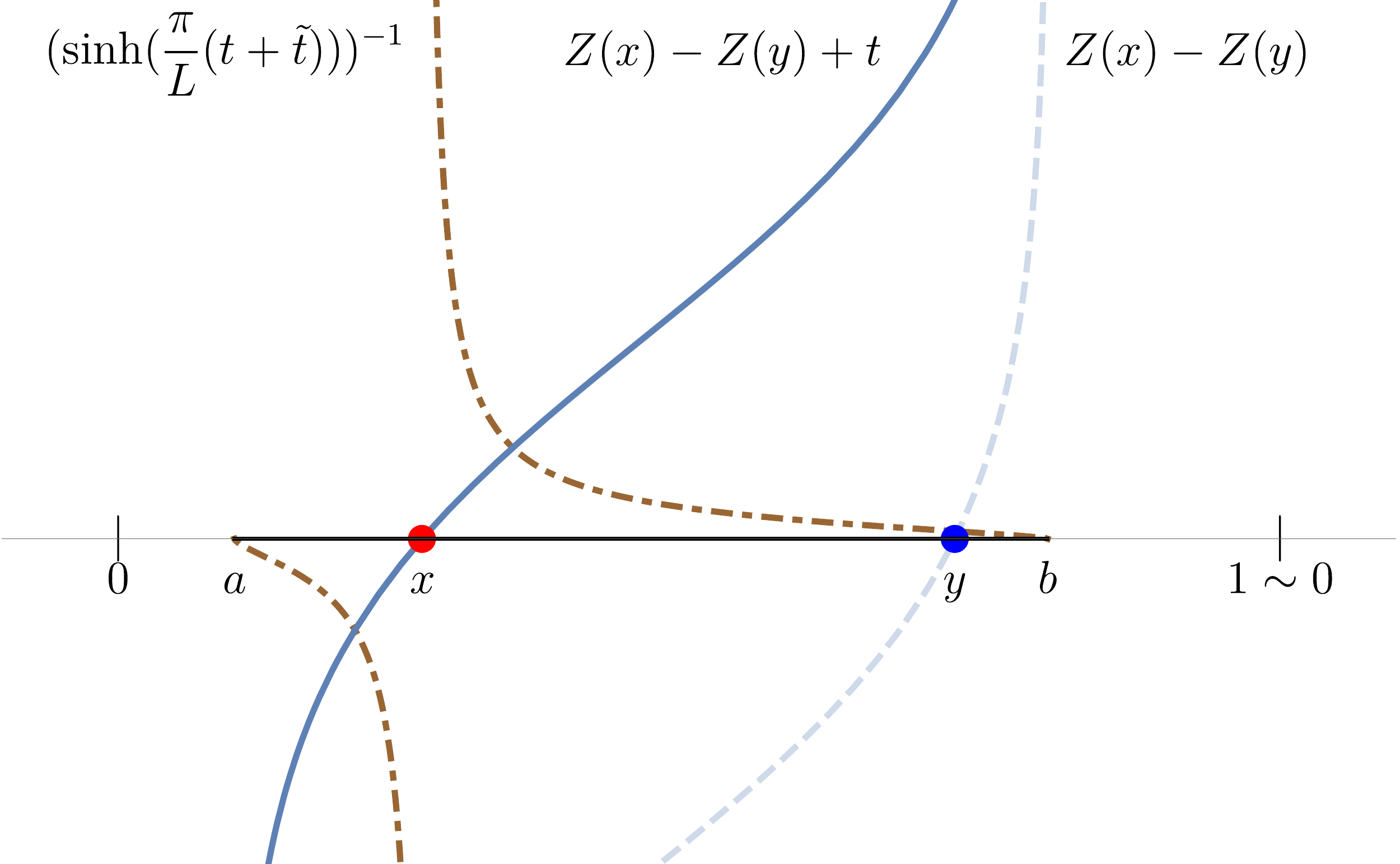}
 \caption{Illustration of the modular flow $\sigma_t\left( \psi^\dag(y) \right)$ of eq. \eqref{sigma_R} for the periodic sector on the cylinder -- one of the novel results of this work -- for a single interval, depicted in a similar manner as figure \ref{fig:two_ints}. Again the blue dot signals the point $y$, while the red dot follows the position of the local contribution to the flow, corresponding to the zero of $Z(x)-Z(y)+t$, for $t=0$ (dashed-opaqued) and $t>0$ (solid blue). The novelty is that in addition, this case contains a completely non-local term, as seen in the second term of \eqref{sigma_R}, represented by the dot-dashed line that is proportional the non-local kernel. This vanishes at the endpoints and diverges at the position of the local contribution. }
 \label{fig:Ramond}
 \end{figure}

The second piece is another important result of this work. It constitutes an example of a \textit{continuously non-local} modular evolution. The operator $\psi^\dag$, initially localised at $y$ at $t=0$, receives contributions from the entire interval as modular time evolves. The properties of the coefficient $\left( \sinh(\pi (t+\tilde t)/L) \right)^{-1}$ are again determined by those of $Z(x)$: since this function increases monotonically and diverges at the endpoints, the coefficient vanishes at $\partial V$, and has an asymptote at the solutions of \eqref{tdZ}. Just as with the bi-local flows of section \ref{subsec:vacuumflow}, the asymptote moves monotonically from $b_i$ at $t\to -\infty$ to $a_i$ at $t\to\infty$. We illustrate these features in figure \ref{fig:Ramond}, for the case of a single interval. 

The corresponding modular Hamiltonian for the periodic case was first derived in \cite{Klich:2015ina}. Although the continuously non-local flow appears to have a different nature than the bi-local quasi-geometric terms, it actually results as the zero temperature limit of bi-locality on the torus, as shown in \cite{Fries:2019ozf}. We comment on this in the next subsection. 

For multiple intervals on the periodic sector, the flow has three components: the local piece, the bi-local term (one per interval as in the plane), and the completely non-local one, which couples a given point to all intervals. 

\subsection{Torus}
The resolvent on the torus possesses a qualitatively new behaviour. As shown in \cite{Fries:2019ozf}, the modular Hamiltonian involves a discrete but infinite set of bi-local couplings. The same behaviour is present for the modular flow and the modular correlator as discussed in section \ref{subsec:mod2pt-torus}. See also \cite{PhysRevD.100.025003} for related work.
 
The novelty here is that the propagator appearing in $F_\lambda$ (see eq. \eqref{eq:f4}) carries an explicit $\lambda$-dependence, and therefore does not factor out of the integral in \eqref{Sigmat2}, as in the previous cases. Nevertheless, we can still find an analytic solution. In \eqref{eq:z4} we noticed that the propagator appearing in this case can be re-interpreted as the correlator with a chemical potential turned on and has an associated series representation, \eqref{eq:gg2} and \eqref{eq:gg3} for each spin sector. Using this, and again essentially the same steps as in section \ref{subsec:vacuumP} for the cylinder (P), one finds
\begin{align}
  \label{eq:sigma_series}
\Sigma^{(\nu)}_t&=\frac{1}{L} \frac{\sinh\left( \pi t \right)}{\sinh \left( \pi \frac{t+\tilde t}{L} \right)}\sum_{k}' \ee^{-2\pi \ii k\left( x-y+\beta (t+\tilde t)/L  \right)}
\end{align}
where the sum runs over $k\in\mathbb Z$ for $\nu=2$ and $k\in\mathbb Z+1/2$ for $\nu=3$. Finally, using the periodic/antiperiodic representations of the Dirac distribution,
\begin{align}
  \label{eq:dirac-comb}
\sum_{k\in \mathbb Z}\ee^{-2\pi \ii k s}=\sum_{k\in\mathbb Z} \delta (s-k)\,,\quad \sum_{k\in \mathbb Z+\frac{1}{2}}\ee^{-2\pi \ii k s}=\sum_{k\in\mathbb Z} (-1)^k\delta (s-k)
\end{align}
we find that the kernel for the modular operator is given, for $\nu=2,3$, by
 \begin{align}\label{}
\Sigma^{(\nu)}_t=\frac{1}{L} \frac{\sinh\left( \pi t \right)}{\sinh \left( \pi \frac{t+\tilde t}{L} \right)} \sum_{k\in \mathbb Z} (-1)^{\nu k}\delta \left( x-y+\beta\frac{t+\tilde t}{L} -k\right)
\end{align}
where we have replaced $\tau=\ii\beta$ again to emphasize that the argument of the Dirac distribution is strictly real. Here again the argument of the distribution plays a fundamental role. The support of $\Sigma_t$ locates at the roots of
\begin{align}\label{tdZ_torus}
x-y+\frac{\beta}{L} \left( t+Z(x)-Z(y) \right) -k=0
\end{align}

\begin{figure}[h]
 \centering
  \includegraphics[width=0.6\textwidth]{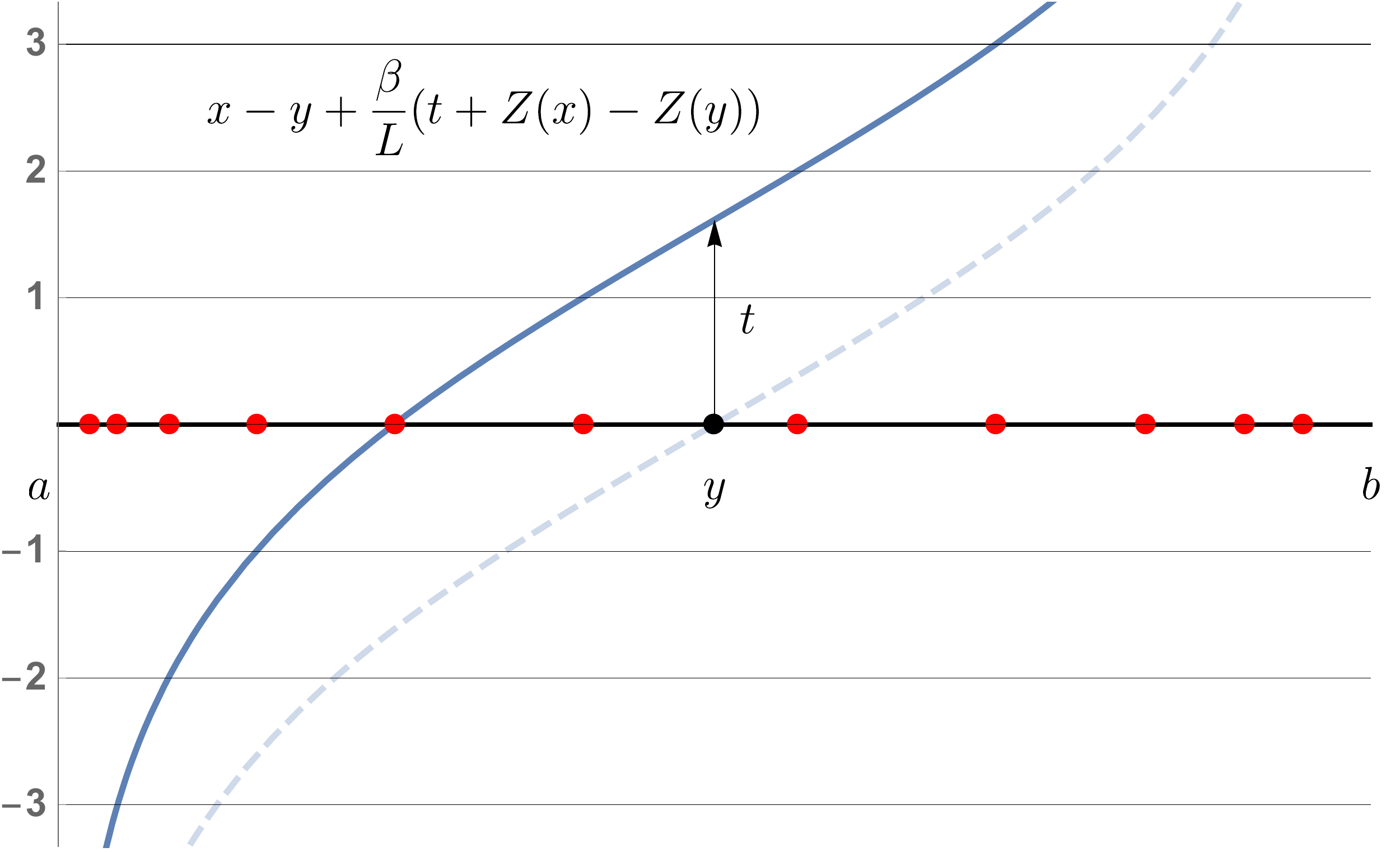}
 \caption{Modular flow \eqref{psi_t_torus} for a single interval on the torus. It mixes an operator at $y$ to an infinite but discrete set of points (red) within the interval. These are the intersections of the solid curve with the integers, solutions to \eqref{tdZ_torus} which accumulate at the endpoints. As modular time increases, these solutions move towards the left endpoint. The dashed opaque curve corresponds to $t=0$, while the solid one is for $t>0$. Note the non-trivial dependence of the curve with the ratio $\beta/L$, the two physical scales of the system. }
 \label{fig:torus}
 \end{figure}

For every integer $k$ and interval $\ell$, \eqref{tdZ_torus} has a single solution, $x_{\ell,k}=x_{\ell,k}(y)$, since again $Z(x)$ is monotonically increasing from $-\infty$ to $+\infty$ within each interval. This is illustrated for a single interval in figure \ref{fig:torus}. Therefore modular flow connects any given point $y\in V$ to an infinite set of other points in $V$. Thus we can write
 \begin{align}\label{}
\Sigma^{(\nu)}_t=\frac{1}{L} \frac{\sinh\left( \pi t \right)}{\sinh \left( \pi \frac{t+\tilde t}{L} \right)} \sum_{k\in \mathbb Z} (-1)^{\nu k} \sum_\ell \frac{\delta(x-x_{\ell,k})}{1+\frac{\beta}{L}Z'(x_{\ell,k})}
\end{align}

Finally, the modular flow on the torus is then
\begin{align}\label{psi_t_torus}
\sigma_t (\psi^\dag(y)) = \frac{\sinh\left( \pi t \right)}{L}  \sum_\ell \sum_{k\in \mathbb Z}   \frac{(-1)^{\nu k}}{\sinh \left(  \frac{\pi}{L}\left( t+Z(x_{\ell,k})-Z(y)  \right) \right)} \frac{\psi^\dag(x_{\ell,k})}{1+\frac{\beta}{L}Z'(x_{\ell,k})}\,.
\end{align}

This result illustrates the `infinite bi-locality', as was already understood for the modular Hamiltonian \cite{Fries:2019ozf,PhysRevD.100.025003}. As can be seen explicitly from the above equation and depicted in figure \ref{fig:torus}, the modular flow of a field localised at $y$ receives contributions from a discrete but infinite set of points within $V$, even for a single interval. As discussed in \cite{Fries:2019ozf}, this structure illuminates the behaviour at zero temperature. Indeed, when $\beta\to\infty$, the solid curve becomes a straight line with infinite slope, and the red dots in figure \ref{fig:torus} form a regular partition of the interval. For $\nu=2$, this reproduces exactly the definition of the Riemann integral, which leads to the continuous integral for the periodic sector on the cylinder as shown in \eqref{sigma_R}.

It is worth emphasizing once more how explicit \eqref{psi_t_torus} is. Indeed, whereas alternative approaches \cite{Casini:2009vk,Blanco:2019cet} use the modular Hamiltonian itself to write the modular flow in terms of a set of infinitely many coupled differential equations, our method yields at once the full solution.    
 	
\section{Modular two-point function}
\label{sec:mod_two}
In this section we illustrate the power of the tools laid out in section~\ref{sec:modular-flows-overview} and explicitly calculate the modular two-point function defined in eq.~\eqref{eq:mod2pt}. Following the distinction of cases from section \ref{subsec:resolvent} we may work out the different results for the plane, the cylinder with (anti-)periodic boundary conditions and finally the torus. We will explicitly calculate $\Gmod(t)$ in the lower strip $-1<\Im(t)<0$, and later comment on the continuation to the complex plane. 

On a formal level, the modular correlator is already determined by the results of the previous section. All we have to do is take the expectation value of $\sigma_t\left( \psi^\dag \right)$ with another fermionic field,
\begin{equation}
	G_\mathrm{mod}(x,y;t)=\left\langle \psi(x)\,\sigma_t(\psi^\dagger(y)) \right\rangle
\end{equation}
on the global state considered, where the modular flow of the operator $\sigma_t(\psi^\dagger(y))$ has been computed for all mentioned cases in section \ref{sec:op_flow}. For instance, in the simplest case of the plane or cylinder (A) this yields
\begin{equation}\label{}
G_\mathrm{mod}(x,y;t)=2\ii \sinh(\pi t) \sum_k \frac{G(x_k,y)G(x,x_k)}{Z'(x_k)}\,,
\end{equation}
where again $x_k(y;t)$ are the solutions of \eqref{tdZ}. In practice however, this formula is not very useful. As mentioned above, the problem of finding the roots $x_k$ is in general very hard. Moreover, as mentioned in section \ref{sec:modular-flows-overview}, the modular correlator must be analytic in the strip $-1\leq\Im(t)\leq 0$ and satisfy the boundary condition~\eqref{eq:kms3}, implied by the KMS condition~\eqref{eq:kms}. Both properties are obscured in the above formula, because it was derived for $t\in\RR$ and the $x_k$ depend implicitly on $t$. Therefore, one would wish for an alternative expression which makes these properties manifest.

Fortunately, such an expression exists. In the next section we compute the modular two-point function directly from the resolvent. Since all the integrals involved are convergent, this yields a final result for the modular correlator that depends on $Z(x)-Z(y)$ rather than on the roots $x_k$. Furthermore, the resulting expression can explicitly be shown to satisfy the boundary condition~\eqref{eq:kms3}.


\subsection{Plane}
\label{subsec:mod2pt-plane}
This is a rather trivial case, but the result is a useful building block for the more involved calculations in the other cases. Note that all results hold for arbitrary configurations of intervals. We want to calculate the modular two-point function \eqref{eq:mod2pt} by solving the variant of Cauchy's integral equation presented in \eqref{eq:cauchy}. After inserting the resolvent given in \eqref{eq:res}, \eqref{eq:f1} and \eqref{eq:z1} we get
\begin{eqnarray}
	G_\mathrm{mod}(x,y;t) & = & \oint_{\gamma}\frac{\dif{\lambda}}{2\pi\ii}\left(\frac{1-\lambda}{\lambda}\right)^{\ii t}\delta(x-y)\\
	&  & -\oint_{\gamma}\frac{\dif{\lambda}}{2\pi\ii}\frac{1}{1-\lambda}\left(\frac{1-\lambda}{\lambda}\right)^{\ii t} G(x,y) \left[-\frac{1-\lambda}{\lambda}\right]^{\ii\tilde{t}}\nonumber\,,
\end{eqnarray}
where we used the shorthand $\tilde{t}=Z(x)-Z(y)$ again and $\gamma$ is a tight counter-clockwise contour around the interval $[0,1]$ as before. The first term vanishes since the integrand is analytic in the integration region. For the second term we find it convenient to substitute $z=(1-\lambda)/\lambda$ and thus we may write
\begin{equation}
	\label{eq:idk}
	G_\mathrm{mod}(x,y;t)=\oint_{\Gamma}\frac{\dif{z}}{2\pi\ii}\frac{z^{\ii t}}{z\left(1+z\right)}\left(-z\right)^{\ii\tilde{t}}G(x,y)\,,
\end{equation}
with $\Gamma$ being a tight counter-clockwise contour around $\RR^+$. We note that $z^{\ii t}$ has a branch cut on $\RR^{-}$ while $\left(-z\right)^{\ii\tilde{t}}$
has a branch cut on $\RR^{+}$ which makes it impossible to
use the residue theorem. However, we can circumvent this problem with a trick that produces a common branch cut on $\RR^{+}$ for the whole integrand. This goes as follows.

\begin{figure}[h]
	\def\svgwidth{0.5\linewidth}
	\centering{
\begingroup%
  \makeatletter%
  \providecommand\color[2][]{%
    \errmessage{(Inkscape) Color is used for the text in Inkscape, but the package 'color.sty' is not loaded}%
    \renewcommand\color[2][]{}%
  }%
  \providecommand\transparent[1]{%
    \errmessage{(Inkscape) Transparency is used (non-zero) for the text in Inkscape, but the package 'transparent.sty' is not loaded}%
    \renewcommand\transparent[1]{}%
  }%
  \providecommand\rotatebox[2]{#2}%
  \newcommand*\fsize{\dimexpr\f@size pt\relax}%
  \newcommand*\lineheight[1]{\fontsize{\fsize}{#1\fsize}\selectfont}%
  \ifx\svgwidth\undefined%
    \setlength{\unitlength}{296.30206075bp}%
    \ifx\svgscale\undefined%
      \relax%
    \else%
      \setlength{\unitlength}{\unitlength * \real{\svgscale}}%
    \fi%
  \else%
    \setlength{\unitlength}{\svgwidth}%
  \fi%
  \global\let\svgwidth\undefined%
  \global\let\svgscale\undefined%
  \makeatother%
  \begin{picture}(1,0.95238268)%
    \lineheight{1}%
    \setlength\tabcolsep{0pt}%
    \put(0,0){\includegraphics[width=\unitlength,page=1]{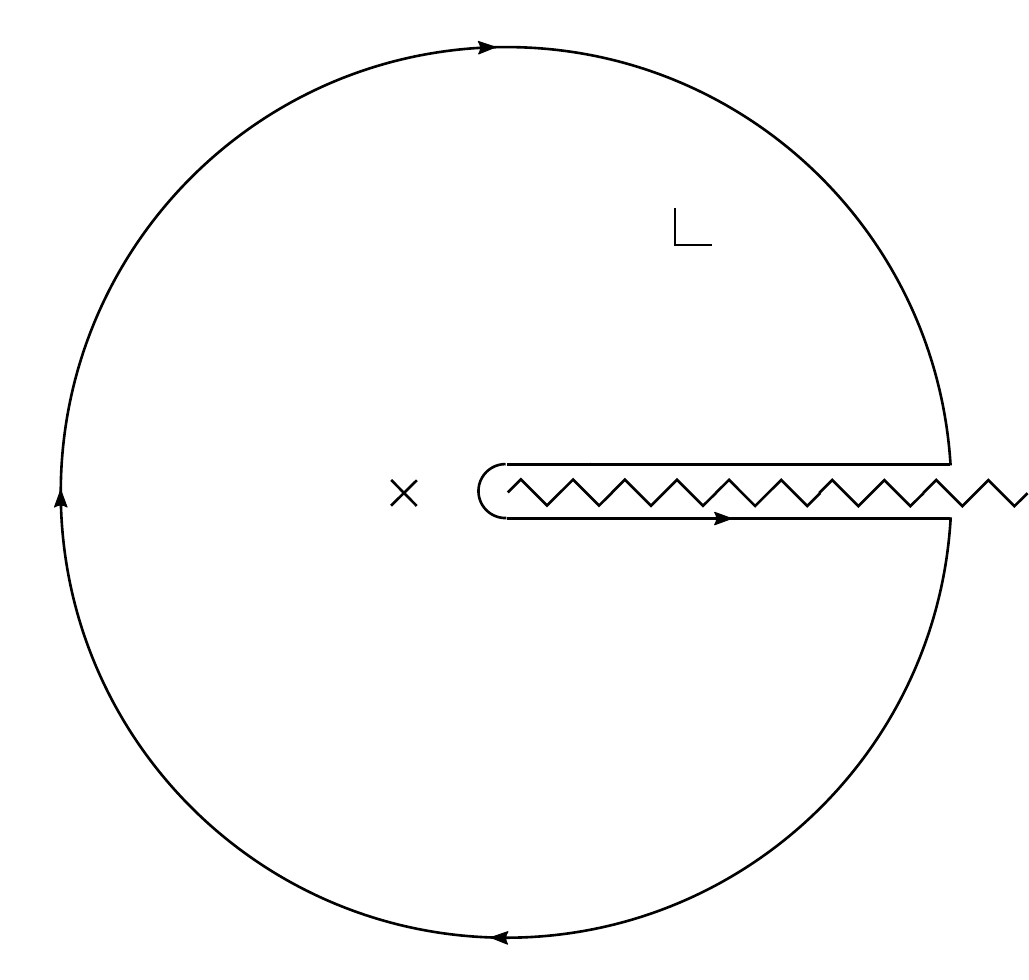}}%
    \put(0.35853059,0.51488751){\color[rgb]{0,0,0}\makebox(0,0)[lt]{\lineheight{1.25}\smash{\begin{tabular}[t]{l}$-1$\end{tabular}}}}%
    \put(0.64677591,0.37695201){\color[rgb]{0,0,0}\makebox(0,0)[lt]{\lineheight{1.25}\smash{\begin{tabular}[t]{l}$\Gamma$\end{tabular}}}}%
    \put(0.66407883,0.72598602){\color[rgb]{0,0,0}\makebox(0,0)[lt]{\lineheight{1.25}\smash{\begin{tabular}[t]{l}$z$\end{tabular}}}}%
    \put(0,0){\includegraphics[width=\unitlength,page=2]{pacman-contour.pdf}}%
  \end{picture}%
\endgroup%

		\caption{The integral in \eqref{eq:pacman} is carried out along the contour $\Gamma$ displayed here. We can close the contour at infinity with a big circle since those contributions vanish. The only non-analyticity in the integration region is the simple pole at $z=-1$ for which we obtain the residue.}
		\label{fig:pacman} 	}
\end{figure}

Assuming $-1<\Im(t)<0$, we can pull tight the integration contour and neglect the contributions at the endpoints $0$ and $\infty$. While $\Im(t)<0$ avoids the divergence at zero, $\Im(t)>-1$ ensures the vanishing of contributions at infinity. Hence, we may write
\begin{eqnarray}
	G_\mathrm{mod}(x,y;t) & = & \int_{0}^{\infty}\frac{\dif{z}}{2\pi\ii}\frac{z^{\ii t}}{z\left(1+z\right)}\left[\left(-z+\ii 0^+\right)^{\ii\tilde{t}}-\left(-z-\ii 0^+\right)^{\ii\tilde{t}}\right]G(x,y)\\
	& = & \int_{0}^{\infty}\frac{\dif{z}}{2\pi\ii}\frac{z^{\ii\left(t+\tilde{t}\right)}}{z\left(1+z\right)}\left[e^{-\pi\tilde{t}}-e^{\pi\tilde{t}}\right]G(x,y)\,.
\end{eqnarray}
Notice that we went from a complex contour integral to an
integral along $\RR^+$ and from $(-z)^{\ii\tilde{t}}$
to $z^{\ii\tilde{t}}$ by acquiring a factor $[e^{-\pi\tilde{t}}-e^{\pi\tilde{t}}]$.
Now we use the inverse logic and go back to the contour integral
and from $z^{\ii(t+\tilde{t})}$ to $(-z)^{\ii(t+\tilde{t})}$
by adding a factor $[e^{-\pi(t+\tilde{t})}-e^{\pi(t+\tilde{t})}]^{-1}$. Thus, we arrive at
\begin{equation}
	\label{eq:pacman}
	G_\mathrm{mod}(x,y;t)= G(x,y)\frac{\sinh\left[\pi\tilde{t}\right]}{\sinh\left[\pi\left(t+\tilde{t}-\ii 0^+\right)\right]}\oint_{\Gamma}\frac{\dif{z}}{2\pi\ii}\frac{\left(-z\right)^{\ii\left(t+\tilde{t}\right)}}{z\left(1+z\right)}\,.
\end{equation}
where we made explicit the above assumption that $\Im(t)<0$. The remaining contour integral is solved by closing $\Gamma$ at infinity with a big circle as illustrated in figure \ref{fig:pacman}. We only pick up the residue at $z=-1$ which is just one. So the final
result for the modular two-point function on the plane is
\begin{equation}
	\label{eq:g-mod12}
	G_\mathrm{mod}(x,y;t)= G(x,y)\frac{\sinh\left[\pi\tilde{t}\left(x,y\right)\right]}{\sinh\left[\pi\left(t+\tilde{t}\left(x,y\right)-\ii 0^+\right)\right]}\,.
\end{equation}

This expression was expected, since it is known already, but it has, to the best of our knowledge, not yet been derived directly from the resolvent. As a first consistency check, we see that for $t=0$ we retain the original two-point function. It is also straightforward to check that it satisfies the KMS condition. Since eq.~\eqref{eq:g-mod12} is analytic for all $t \in \CC$, we do not have to care about the direction of any limits and can straighforwardly substitute $t\rightarrow t-\ii$. Then, the $\sinh$ in the denominator switches sign and thus the entire expression switches sign. This verifies KMS condition in the form of~\eqref{eq:kms3} since the right hand side vanishes up to the localized contribution~\eqref{Sigmasol}.

\subsection{Cylinder}
\label{subsec:mod2pt-cylinder}
Now we will take the spatial direction to be periodic with periodicity 1. For the boundary conditions there are two possibilities. In the first case of antiperiodic boundary conditions, there is no more work to be done. Recalling the details given in section~\ref{subsec:resolvent} we see that the only difference to the case of the plane lies in the definitions of the propagator \eqref{eq:g2} and the function $Z(x)$ \eqref{eq:z23}. However, both of these definitions did not play a role in the calculation. Hence, the modular two-point function on the cylinder with antiperiodic boundary conditions is already given by the expression in \eqref{eq:g-mod12}.

The case of periodic boundary conditions is more complicated insofar as the resolvent \eqref{eq:f3} features an additional term. The corresponding contribution to the modular two-point function is
\begin{eqnarray}
	\delta G_\mathrm{mod}(x,y;t) & = &  -\oint_{\gamma}\frac{\dif\lambda}{2\pi\ii}\frac{1}{1-\lambda}\left(\frac{1-\lambda}{\lambda}\right)^{\ii t}\left(\frac{\lambda-1}{\lambda}\right)^{\ii\tilde{t}}\frac{1}{2}\frac{\left(\frac{\lambda-1}{\lambda}\right)^{L}-1}{\left(\frac{\lambda-1}{\lambda}\right)^{L}+1}\\
	& = & \oint_{\Gamma}\frac{\dif{z}}{2\pi\ii}\frac{1}{z\left(1+z\right)}z^{\ii t}\left(-z\right)^{\ii\tilde{t}}\frac{1}{2}\frac{\left(-z\right)^{L}-1}{\left(-z\right)^{L}+1}\,,
\end{eqnarray}
where we used the same definitions for $z$, $\tilde{t}$, $\gamma$ and $\Gamma$ as in the previous subsection. We may split the expression into two terms
\begin{eqnarray}
	\delta G_\mathrm{mod}(x,y;t) & = & \frac{1}{2}\oint_{\Gamma}\frac{\dif{z}}{2\pi\ii}\frac{1}{z\left(1+z\right)}z^{\ii t}\left(-z\right)^{\ii\tilde{t}} \label{eq:club}\\
	&  & -\oint_{\Gamma}\frac{\dif{z}}{2\pi\ii}\frac{1}{z\left(1+z\right)}\frac{z^{\ii t}\left(-z\right)^{\ii\tilde{t}}}{\left(-z\right)^{L}+1}\nonumber
\end{eqnarray}
after which we recognize the first term to be the same integral as in \eqref{eq:idk} whose solution has been given in \eqref{eq:g-mod12}. Let us refer to the second term as $\Delta G_\mathrm{mod}$. To solve it we make use of the identity \eqref{eq:identity}. It follows that
\begin{equation}
	\Delta G_\mathrm{mod}=-\frac{\ii}{2}\int_{-\infty}^{\infty}\frac{\dif{s}}{\sinh\left(\pi s+\ii 0^+\right)}\oint_{\Gamma}\frac{\dif{z}}{2\pi\ii}\frac{1}{z\left(1+z\right)}z^{\ii t}\left(-z\right)^{\ii\left(\tilde{t}+sL\right)}\,.
\end{equation}
Again, the contour integral is of the same type as in \eqref{eq:idk}, which leaves us with the integral over $s$. We find it convenient to substitute $w=\ee^{\pi s}$ and arrive at
\begin{equation}
	\label{eq:spade}
	\Delta G_\mathrm{mod}=\frac{\ee^{-\pi t}}{\pi\ii}\int_{0}^{\infty}\frac{\dif{w}}{w^{2}-1+\ii 0^+}\cdot\frac{w^{2L}-\ee^{-2\pi\tilde{t}}}{w^{2L}-\ee^{-2\pi\left(t+\tilde{t}\right)}}\,.
\end{equation}

Our further solution is restricted to the special case $L=\nicefrac{1}{n},n\in\NN\setminus\{1\}$. Under this restriction we are able to separate the integrand into partial fractions which results in a sum of simple integrals. To do that we continue with the substitution $v=w^{\nicefrac{1}{n}}$, yielding
\begin{equation}
	\Delta G_\mathrm{mod}^{L=\nicefrac{1}{n}}=\frac{n\,\ee^{-\pi t}}{\pi\,\ii}\int_{0}^{\infty}\dif{v}\frac{v^{n-1}\left(v^{2}-\ee^{-2\pi\tilde{ t}}\right)}{\left(v^{2n}-1+\ii 0^+\right)\left(v^{2}-\ee^{-2\pi\left( t+\tilde{ t}\right)}\right)}\,.
\end{equation}
A decomposition into partial fractions gives
\begin{eqnarray}
	\label{eq:partial-frac}
	\Delta G_\mathrm{mod}^{L=\nicefrac{1}{n}} & = & \frac{1}{2\pi\ii}\int_{0}^{\infty}\dif{v}\left\{ \frac{\sinh\left[\pi\tilde{ t}\right]}{\sinh\left[\pi\left( t+\tilde{ t}\right)\right]}\cdot\frac{1}{v-1+\ii 0^+}\right.\\
	&  & +\sum_{k=1}^{2n-1}\left(-1\right)^{k}\frac{\sinh\left[\pi\left(\tilde{ t}+\ii\frac{k}{n}\right)\right]}{\sinh\left[\pi\left( t+\tilde{ t}+\ii\frac{k}{n}\right)\right]}\cdot\frac{1}{v-\ee^{i\pi\frac{k}{n}}}\nonumber \\
	&  & +\left.n\frac{\sinh\left[\pi t\right]}{\sinh\left[n\,\pi\left(t+\tilde{t}\right)\right]}\left[\frac{1}{v-\ee^{-\pi\left(t+\tilde{t}\right)}-\ii 0^+}+\frac{\left(-1\right)^{n-2}}{v+\ee^{-\pi\left( t+\tilde{ t}\right)}}\right]\right\}\,.\nonumber
\end{eqnarray}
In the third line we made the negative imaginary part of $t$ explicit
as a prescription for how to avoid the pole. Although all terms diverge separately, we can trust the divergences to cancel in total, since the
original integral is finite. Importantly, note that the boundary term of the first summand cancels exactly with the first term of $\delta G_\mathrm{mod}$ in \eqref{eq:club}. Thus, our final result for the modular two-point function on the cylinder with periodic boundary conditions is
\begin{equation}
	\label{eq:g-mod-cyl}
	G_\mathrm{mod}^{L=\nicefrac{1}{n}}(x,y;t)=
	G(x,y)\frac{\sinh\left[\pi\tilde{t}\right]}{\sinh\left[\pi\left(t+\tilde{t}\right)\right]}
	+\delta G_\mathrm{mod}^{L=\nicefrac{1}{n}}(x,y;t)
\end{equation}
with
\begin{eqnarray}
	\label{eq:g-mod3}
	\delta G_\mathrm{mod}^{L=\nicefrac{1}{n}}(x,y;t) & = & \frac{1}{2}\sum_{k=1}^{2n-1}\left(1-\frac{k}{n}\right)\left(-1\right)^{k}\frac{\sinh\left[\pi\left(\tilde{ t}+\ii\frac{k}{n}\right)\right]}{\sinh\left[\pi\left( t+\tilde{ t}+\ii\frac{k}{n}\right)\right]}\\
	&  & +\frac{n}{2}\frac{\sinh\left[\pi t\right]}{\sinh\left[n\,\pi\left( t+\tilde{ t}\right)\right]}\left[1-\left(1+\left(-1\right)^{n-2}\right)\ii\left( t+\tilde{ t}\right)\right]\,.\nonumber
\end{eqnarray}

It is straightforward to check that $\delta G_\mathrm{mod}$ vanishes for $t=0$. This is as expected, since we should recover the usual propagator. To see what happens to the KMS condition, we again insert $t\rightarrow t-\ii$. The first term clearly switches sign hence does not contribute to the left hand side of eq.~\eqref{eq:kms3}. For the second term, we can to consider the cases of odd and even $n$ separately. In both cases, we end up with
\begin{align}\label{eq:jump}
  \delta G_\mathrm{mod}^{L=\nicefrac{1}{n}}(x,y;t-\ii 0^+) + \delta G_\mathrm{mod}^{L=\nicefrac{1}{n}}(x,y;t-\ii+\ii 0^+)= \frac{n \sinh\left[\pi t\right]}{\sinh\left[n\,\pi\left( t+\tilde{ t}\right)\right]},
\end{align}
reproducing exactly the non-local extra term from~\eqref{sigma_R}. As far as the authors know, this is the first time in the literature that such a non-local term was explicitly derived in the context of modular two-point functions.

\subsection{Torus}
\label{subsec:mod2pt-torus}
At last, we will turn to the modular two-point function on the torus, i.e. we will deal with periodicity 1 in the spatial direction and with periodicity $\tau=\ii\beta$ in the time direction. Following the labelling in section~\ref{subsec:resolvent}, the boundary conditions will be denoted by $\nu=2,3$. To begin with, the same recipe as in the previous subsections can be applied. We take the necessary information about the resolvent and the propagator from the third key point in section~\ref{subsec:resolvent}. It strikes us that we can (and will) make use of \eqref{eq:identity} once again.

Let us demonstrate the procedure for $\nu=2$ and then comment on the difference for $\nu=3$. Combining all things mentioned (again with $z$, $\tilde{t}$ and $\Gamma$ defined as before) we have
\begin{equation}
G_\mathrm{mod}(x,y;t)=\frac{\ii}{2}\int_{-\infty}^{\infty}\dif{s}\frac{\sum_{k\in\ZZ}'\ee^{-2\pi\ii k(x-y-\ii 0^++\beta s)}}{\sinh\left(\pi s+\ii 0^+\right)}\oint_{\Gamma}\frac{\dif{z}}{2\pi\ii}\frac{z^{\ii t}}{z\left(1+z\right)}\left(-z\right)^{\ii\left(\tilde{t}-Ls\right)}\,,
\end{equation}
which allows a number of simplifications. The contour integral over $z$ on the right might be an old acquaintance by now. It is of the same type as the one in \eqref{eq:idk}, which was solved by \eqref{eq:g-mod12}. The sum over $k$ on the left can be recognized as a Dirac comb with periodicity~1. Putting the changes into place results in
\begin{eqnarray}
G_\mathrm{mod}(x,y;t) & = & 
\frac{\ii}{2\beta}\int_{-\infty}^{\infty}\dif{s}\frac{\sum_{n\in\ZZ}'\delta\left(s+\frac{x-y-n}{\beta}\right)}{\sinh\left(\pi s+\ii 0^+\right)}\cdot\frac{\sinh\left[\pi\left(\tilde{t}-Ls\right)\right]}{\sinh\left[\pi\left(t+\tilde{t}-Ls\right)\right]} \\
& = & \frac{\ii}{2\beta}\sum_{n\in\ZZ}'\frac{1}{\sinh\left(-\pi\frac{x-y-n}{\beta}+\ii 0^+\right)}\cdot\frac{\sinh\left[\pi\left(\tilde{t}+L\frac{x-y-n}{\beta}\right)\right]}{\sinh\left[\pi\left(t+\tilde{t}+L\frac{x-y-n}{\beta}\right)\right]}\,, \label{eq:g-mod-torus}
\end{eqnarray}
which is our final result for the modular two-point function on the torus with PA boundary conditions ($\nu=2$). To see the slight difference for AA boundary conditions ($\nu=3$), it suffices to compare eqs.~\eqref{eq:gg2} and \eqref{eq:gg3}. The sum is now over $k\in\ZZ+1/2$ which produces an additional factor $(-1)^n$ in the final result.

For $t=0$, the second factor in \eqref{eq:g-mod-torus} cancels and we can trace back the steps to $G(x,y)$ as expected. To check for the KMS condition, we replace $t\rightarrow t-\ii$ and confirm that the expected sign change is produced by the $\sinh$ in the denominator of the second factor, up to the isolated poles which yield local contributions on the right hand side of~\eqref{eq:kms3}.

\subsection{Analytic structure}
\label{subsec:mod2pt-poles}

As a key feature of two-point functions we deem it instructive to analyze the pole structure of the above results. This will give insights into the non-locality of modular flow, since it causes couplings between multiple points or even entire regions. 

We illustrate the analytic structure of $\Gmod(x,y,t)$ for $t\in\mathbb C$, for some fixed $x,y\in V$. As anticipated in section \ref{sec:modular-flows-overview}, the presence of poles and cuts gives information about the locality or not of the modular evolution. First, we know that $\Gmod(x,y,t)$ possesses simple poles along the real axis, located at the solutions to $\eqref{tdZ}$ and \eqref{tdZ_torus} for the plane/cylinder, and the torus respectively. Due to the definition of the modular correlator and the KMS condition \eqref{eq:kms2}, we also know that the function is antiperiodic in imaginary modular time. The precise values of $x,y$ -- and therefore the poles -- will not be important for the discussion.  

\paragraph{Plane or cylinder (A).} The modular two-point function in the plane or on the cylinder with antiperiodic boundary conditions was determined in \eqref{eq:g-mod12}. Given fixed $x,y$, the poles are the values of $t$ which satisfy
\begin{equation}\label{eq:sinhtdZ}
\sinh \left(  t+Z(x)-Z(y) \right)=0\,. 
\end{equation}

\begin{figure}[h]
	\def\svgwidth{.7\linewidth}
	\centering{
		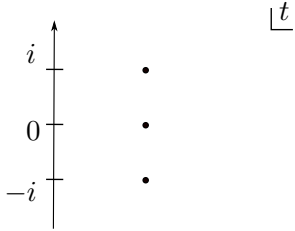
		\caption{ Sketch of the analytic structure of the modular correlator $\Gmod(x,y,t)$ \eqref{eq:g-mod12} as function of complex modular time $t$, for fixed $x,y$ belonging to a single interval on the plane or cylinder(A). The function is analytic everywhere except at its simple poles (black dots), their specific location depending on $x,y$.  Since the modular flow in this case is local, $\Gmod(t)$ possesses no branch cuts. For multiple intervals, the location of the poles is shifted, but no new poles arise. The KMS condition ensures that $\Gmod$ is antiperiodic in imaginary time. }
                \label{fig:Gmod-plane} 	}
\end{figure}

Notice that, since $Z(x)\in\mathbb R$ for $x\in V$ (and similarily for $y$), the solutions of this equation lie at $\Im(t)\in \mathbb Z$. In figure \ref{fig:Gmod-plane} we illustrate the structure of the modular two-point function when continued to the complex plane. The sketch can be easily understood with the aim of figure \ref{fig:two_ints}. As portrayed there, if we fix $y$ and evolve in $t$, the values of $x$ that satisfy \eqref{eq:sinhtdZ} move, and sweep the entirety of $V$ when $t$ varies in $(-\infty,\infty)$. In other words, for fixed $x,y\in V$ there exists a unique value of $t\in\mathbb R$ that solves \eqref{eq:sinhtdZ}. This is a poles of $\Gmod$, depicted as a black dot in figure \ref{fig:Gmod-plane}, which gets repeated due to antiperiodicity.

\begin{figure}[h]
	\def\svgwidth{.7\linewidth}
	\centering{
		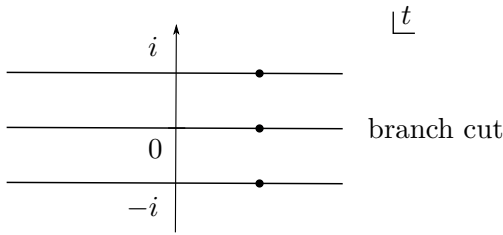
		\caption{ Sketch of $\Gmod(x,y,t)$ \eqref{eq:g-mod-cyl}, for fixed $x,y$ belonging to a single interval on the plane or cylinder (P). The function is analytic in the interior of each strip and antiperiodic amongst strips due to the KMS condition. Just as in the plane, it possesses simple poles (black dots) along $\Im(t)\in \mathbb Z$. However, the novelty here is that in addition it has branch cuts, indicating that the modular flow is completely non-local as explained around \eqref{eq:kms3}. }
                \label{fig:Gmod-R} 	}
\end{figure}

\paragraph{Cylinder (P).} On the cylinder with periodic boundary conditions we have the same situation as above plus the contributions coming from the additional term $\delta \Gmod$ in  \eqref{eq:g-mod3}. This is the novel ingredient in this result. As explained in \eqref{eq:jump}, this contribution induces a discontinuity along $\Im(t)\in\mathbb Z$, as depicted in figure \ref{fig:Gmod-R}. At first sight, the appearance of a cut might seem surprising, due to its absence in the antiperiodic sector. However, this will be clarified in brief once we discuss the case on the torus, and consider the limit of zero temperature. 

As a remark, notice that there is no cut at $t\in i\mathbb Z$, since by definition the modular two-point function coincides with the propagator at $t=0$. 

\paragraph{Torus.} To analyze the pole structure of the modular two-point function for the torus with PA boundary conditions, we reconsider our result \eqref{eq:g-mod-torus}. The first factor inside the sum, although it seems obscure, represents the usual propagator $G(x,y)$ and is not important for our discussion. The second factor determines the interesting poles as the solutions to
\begin{equation}
	\sinh\left( Z(x)-Z(y)-\frac{L}{\beta}\left( y-x+k \right)+t \right)=0\ \ \ \ ,\ \ \ k\in\mathbb Z\,.
\end{equation}
 Since we already know the effect that a composition of multiple disjoint intervals has, let us focus on the case of a single interval. Even then there are countably infinitely many solutions for $t$, namely one for each $k$, spaced by $L/\beta$. Hence, the modular two-point function on the torus (PA) shows a non-locality that consists of infinitely many bi-local couplings.

\begin{figure}[h]
	\def\svgwidth{.7\linewidth}
	\centering{
		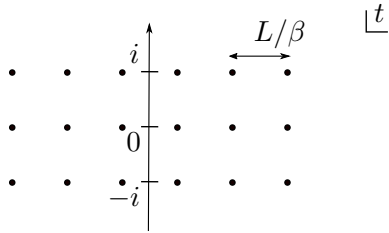
		\caption{ Sketch of the analytic structure of the modular correlator $\Gmod(x,y,t)$ \eqref{eq:g-mod-torus} for fixed $x,y$ belonging to a single interval on the torus. The function is analytic everywhere except at simple poles (black dots), located at the solutions to \eqref{tdZ_torus}, which lie on a lattice of spacings $i$ and $L/\beta$ respectively. In the limit of low temperature $\beta\to\infty$, the real component of the spacing between poles vanishes. This behaviour is the origin of the branch cut in the case of the cylinder (P), figure \ref{fig:Gmod-R}.  }
                \label{fig:Gmod-torus} 	}
\end{figure}

In the low-temperature limit $\beta/L\rightarrow\infty$ we note that the poles move closer to each other until they finally form a branch cut, precisely the situation that we found for the cylinder (P). This is consistent with the interpretation of the cylinder as the low-temperature limit of the torus. By the same line of argument, we expect this infinite set of poles not to appear for the torus with AA boundary conditions. Indeed, the factor $(-1)^n$ in the sum of \eqref{eq:g-mod-torus} causes the poles to cancel each other, in a way completely analogous to the modular Hamiltonian \cite{Fries:2019ozf}.

In order to understand how a cut might originate from an infinite set of simple poles, consider the following example: 
\begin{equation}
	f(x)=\int_{a}^{b} \dif{\lambda}\,\frac{1}{x-\lambda}=\log\left(\frac{a-x}{b-x}\right)\,.
\end{equation}
Here, we produced a function with a branch cut in the interval $[a,b]$ by `summing' over infinitely many simple poles between $a$ and $b$. This is precisely what happens in the low temperature limit of the torus, where the lattice of simple poles create a branch cut in the modular two-point function. In turn, this implies that for any fixed modular time $t$, the modular flow on the cylinder couples a given point $y$ to the entirety of the region $V$.   


\section{Discussion}
\label{sec:conclusions}

In this paper we computed the modular flow for the chiral fermion CFT in $1+1$ dimensions,
for entangling regions consisting of arbitrary sets of disjoint intervals. Working in the framework of functional calculus, we derived two important formulae: 
1) The modular flow of the field operators \eqref{tomitapsi}, 2) the associated modular two-point
function \eqref{eq:mod2pt}. This was done for the cases of the vacuum and thermal states, both on
the infinite line and on the circle, giving an extensive overview of cases that are
usually of interest\,\cite{DiFrancesco:1997nk}. 

A central element in our analysis is that we made extensive use of the resolvent method. 
This technique has allowed us to resolve two main obstacles in the understanding of fermionic entanglement. First, we computed modular flow directly, bypassing the need for the modular Hamiltonian. This is important because although the modular Hamiltonian for the cases of interest was known\,\cite{Casini:2009vk,Klich:2015ina,Fries:2019ozf,PhysRevD.100.025003}, determining the flow from the Hamiltonian is in general very involved and remained unknown. The second problem one faces is that, even with the knowledge of the operator flow, finding the modular two-point function analytically appears hopeless at first sight. The reason is that, as discussed in section \ref{sec:mod_two}, the operator flow generically involves solving transcendental equations whose solutions are unknown. However, we have shown that the resolvent yields directly the modular correlator in closed analytic form, without even the need of considering such equations.  

On the circle, we considered both antiperiodic (A) and periodic (P) boundary conditions, yielding
closed form expressions for both modular flows, including the extra terms that appear due to a zero mode
contribution in the periodic sector. These extra terms appear in the modular two-point function as a branch
cut --- a feature which, to our knowledge, is novel for exact solutions and which can be
traced back to the non-locality of the associated modular flow, as we discussed in section \ref{subsec:mod2pt-poles}. On the torus, we considered the $\nu=2,3$ sectors, showing that the flow leads to an infinite bi-local set of couplings. Moreover, we illustrated how the analytic structure of the two vacua on the cylinder arises as the low temperature limit of the exact result on the torus. 


These solutions display a spectrum of degrees of non-locality. To discuss them, let us
restrict to the periodic case: Starting at high temperature, the correlations are
dominated by thermal fluctuations. Entanglement is thus suppressed and the reduced density
matrix is still a thermal state, albeit with a different temperature $\beta/L$ due to
the introduction of the region size $L$ as an additional scale. As a result, modular flow
coincides with regular, local, time evolution. While this is to be expected in fairly
general theories, it can also be seen explicitly from our results in
eq.~\eqref{tdZ_torus},\eqref{psi_t_torus} and figure~\ref{fig:torus}: In the limit of
small $\beta$, the $Z$-terms vanish, and a field initially localized at $y$ is
transported to the point $x = y - t\beta/L$. As we lower the temperature, we see that
modular flow adds bi-local couplings between an infinite discrete set of points, one for
each value of $k$ in eq.~\eqref{tdZ_torus}. As we derived in
eq.~\eqref{eq:sigma_series},\eqref{eq:dirac-comb}, these additional couplings can be
traced back to the anti-periodicty of the thermal propagator, i.e., the KMS condition in
physical time. Finally, as the temperature approaches absolute zero, the discrete
couplings condense to a continuum, which we derived in eq.~\eqref{sigma_R}.

As we saw in this paper, we have explicit results for mixed states in regions that consists of arbitrary sets of disjoint intervals. We cover finite temperature states, as well as that of the perfect mixture of degenerate ground states. It is remarkable how far one can actually go: This method provides not only a recipe to determine modular flow ‘in principle’, but — by virtue of the power of complex analysis — allows to determine modular flow in closed analytic form. 

Let us point out some future directions. In addition to naturally continuing the work on the modular theory for free fermions on
the torus, our results may give a starting point for computations of more
general modular flows. For example, the methods presented in this paper can readily be applied to excited states in the
fermion CFT, since the corresponding resolvent is known \cite{Klich:2015ina}. We 
expect to find additional non-local terms there. 

Another possible direction for further progress lies in the generalization to higher
dimensions, and even to massive (i.e.~non-conformal)  theories: Since the resolvent formulae that we derived are valid
for \emph{any} free fermionic theory, they can be readily applied if one has the
corresponding resolvent at hand, which in turn can always be found by solving an integral
equation.

Finally, since the algebra of higher spin fields embeds into
the free fermion CFT \cite{Bischoff:2011mx}, it will be interesting to see if similar
methods can be used to derive modular flows there.  We expect this to have applications to higher
spin AdS/CFT\,\cite{Campoleoni:2010zq,Gaberdiel:2010pz,Castro:2011zq,Ammon:2013hba,deBoer:2013vca} and to bulk reconstruction in this context. A starting point in this direction will be to  explore the modular flow of higher spin operators by adapting the techniques presented here.  

\bigskip

{\bf Acknowledgements}

We are grateful to Haye Hinrichsen for discussions.
The work of IR is funded by the Gravity, Quantum Fields and Information group at AEI, which is generously supported by the Alexander von Humboldt Foundation and the Federal Ministry for Education and Research through the Sofja Kovalevskaja Award. PF is  supported by the DFG project HI 744/9-1. 


\appendix

\section{Regularisation of the modular operator}
\label{appendixA}

In this section, we provide a more detailed presentation of the regularisation involved in defining the modular operator via the resolvent, i.e. \eqref{Sigmat1}. We shall consider the case of the plane, but the generalisation to the other cases is straightforward. After replacing the resolvent \eqref{eq:f1} and changing to the variable $z=(1-\lambda)/\lambda$, the operator to be computed is
\begin{align}\label{}
\Sigma_t=-\frac{1}{2\pi i}\oint_\Gamma \frac{dz}{z}z^{it} \left[ \frac{z}{1+z}-(-z)^{i\tilde t}G \right]\,,
\end{align}
where the contour $\Gamma$ is depicted in \ref{fig:cont_3}. This integral is not well defined, as the contributions at both the origin and infinity are not properly defined. Nevertheless, it is easy to perform a regularisation to give it a precise meaning as follows. In order to tame both the origin and infinity, consider instead the function \cite{Fries:2019ozf}
\begin{align}\label{eq:Sigmae}
\Sigma^{(\epsilon)}_t&=-\frac{1}{2\pi i}\oint_\Gamma \frac{dz}{z}z^{it} \left[ \frac{z}{1+z}-G(-z)^{i\tilde t} \right] \cdot \frac{\left( 1-\epsilon^{-1} \right)z}{(z-\epsilon)\left( z-\epsilon^{-1} \right)}\,.
\end{align}

The last fraction contains the regulating function. When $z\to 0$, this vanishes linearly, which cancels the pole at the origin rendering the integrand bounded there. As $z\to \infty$ it decays linearly, which provides the necessary falloff for the contribution at infinity to converge. Finally, we see that
\begin{align}\label{}
\Sigma_t=\lim_{\epsilon\to 0} \Sigma_t^{(\epsilon)}\,.
\end{align}

The regularised integral $\Sigma_t^{(\epsilon)}$ is now easily computed. We have
\begin{align}\label{}
\Sigma^{(\epsilon)}_t&=-\frac{1}{2\pi i}(1-\epsilon^{-1})\oint_{\Gamma} \frac{dz}{z-\epsilon}z^{it} \left[ \frac{z}{1+z}-G(-z)^{i\tilde t} \right]\frac{1}{z-\epsilon^{-1}}\,,
\end{align}
which possesses two single poles, at $z=\epsilon$ and $z=\epsilon^{-1}$ respectively. The first term in the integral vanishes, because the function is holomorphic in a small neighbourhood around $\Gamma$ which does not contain any of the poles. This explains the transition from \eqref{Sigmat1} to \eqref{Sigmat2} in the main text.  This is consistent with the resolvent method, because with our approach, all the contribution should come from the branch cut of the resolvent which is contained in the second term. Thus we are left with
\begin{align}\label{eq:Sigmae2}
\Sigma^{(\epsilon)}_t=\frac{1}{2\pi i}(1-\epsilon^{-1})G\oint_{\Gamma} dz\frac{z^{it} (-z)^{i\tilde t} }{(z-\epsilon)(z-\epsilon^{-1})}\,.
\end{align}

Now the discontinuity along $\mathbb R^+$ implies that just above and below the cut we have
\begin{align}\label{}
\left( -(z\pm i0^+) \right)^{i\tilde t}=e^{\pm \pi \tilde t} z^{i\tilde t}\,.
\end{align}
Again, since after regularization the integrand is bounded in a vicinity of the origin, there is no contribution around the origin, and we are left with
\begin{align}\label{eq:Sigmae2}
\Sigma^{(\epsilon)}_t=\frac{1}{2\pi i}(1-\epsilon^{-1})G \left( e^{\pi\tilde t}-e^{-\pi\tilde t} \right)\int_0^\infty dz\frac{z^{i(t+\tilde t)} }{(z-\epsilon)(z-\epsilon^{-1})}\,.
\end{align}
This last integral is readily evaluated. Relabelling the regulator as $\epsilon=-ie^{-2\pi m}$, in the limit $m\to\infty$ it yields \eqref{Sigmasol}.

\bibliographystyle{utphys}
\bibliography{biblio}

\end{document}